\documentclass[12pt]{article}
\usepackage{latexsym}
\usepackage{graphics}
\usepackage{epsfig}
\usepackage{booktabs}
\usepackage{epstopdf}
\usepackage{amsmath}
\usepackage{cite}
\usepackage{hyperref}
\usepackage[toc,page]{appendix}
\usepackage{amssymb}

\begin{document}

\begin{center}
\textbf{\Large 
Cluster analysis of earthquake hypocenters in Azerbaijan and surrounding territories
} \vspace{0.5 cm}

Sergii Skurativskyi$^1$\footnote{e-mail: \url{skurserg@gmail.com}},  Sergiy Mykulyak$^1$ \footnote{e-mail: \url{mykulyak@ukr.net}},  Yuliya Semenova $^{1,2}$ \footnote{e-mail: \url{semenova.igph@gmail.com}},   Kateryna Skurativska $^3$  \vspace{0.5 cm} \footnote{e-mail: \url{kateryna.skur@gmail.com}}

$^1$ S.I.Subbotin Institute of Geophysics of the National Academy of Science of Ukraine, Kyiv, Ukraine

 $^2$ DGFI-TUM,  Technical University of Munich, Munich, Germany

$^3$    University of Padova, Padova, Italy 
\end{center}

\begin{quote} \textbf{Abstract.}{\small 
The research focuses on seismic events that occurred in Azerbaijan and adjacent territories, regions known for strong seismic activity. We analyze a catalog of recorded earthquakes between 2010 and 2023, extracting the locations of the earthquake hypocenters for study purposes. Using statistical methods and cluster analysis tools, we developed a procedure for partitioning hypocenter clusters.
The procedure begins with estimates of the Morisita Index, which is suitable for preliminary assessments of the statistical properties of hypocenter sets. Analysis of the Morisita Index indicates that the spatial distribution of hypocenters is heterogeneous, containing denser domains referred to as clusters.
The next stage involves identifying spatial clusters using the DBSCAN and HDBSCAN algorithms. Due to the strong dependence of results on the algorithm's parameters, we selected several partitions with 5–8 clusters that provided maximal or near-maximal Silhouette Index values. The final stage assesses the similarity of the resulting partitions, using the Adjusted Rand Index to identify partitions with a specified degree of similarity.
The final set of partitions was compared to the fault network of the region. Based on the selected partition, the earthquake depth distributions were studied. Specifically, approximate probability density functions were constructed in the form of mixtures of normal distributions, leading to the identification of several bimodal distributions.
}
\end{quote}

\begin{quote} \textbf{Keyword:}{\small 
Morisita index, Azerbaijan seismicity,  focal depth distribution
}
\end{quote}

\vspace{0.5 cm}

\section*{Introduction}
%1,orcid.org/0000-0001-7282-5886
%I.A.Skurativska *1,
%orcid.org/0000-0001-7129-4980
%1,
%orcid.org/0009-0006-8836-2824
%O.M. Sizonenko 2,
%orcid.org/0000-0002-8449-2481
%I.M. Hubar 1,
%orcid.org/0000-0002-2822-7288

\section{Introduction}

The Caucasus is a region where the Arabian and Eurasian plates collide intensely. As highlighted by \cite{Tsereteli_faults, opentech, ISMAILZADEH2020}, this area is highly tectonically active, marked by the formation of seismogenic faults, fold-and-thrust belts, and depressions that are indicative of ongoing orogenic processes.

This tectonic activity gives rise to seismically active zones, characterized by a significant number of seismic events, including strong earthquakes with magnitudes reaching up to $M = 7$ \cite{ISMAILZADEH2020}. Despite the high seismic hazard in these regions, they are often densely populated, host critical infrastructure, and support economic activities, including the extraction of mineral and agricultural raw materials.
 
To comprehend the evolutionary scenarios of this geosystem, secure the sustainable functioning of the regional community, and develop strategies to mitigate seismic hazards, comprehensive studies of the Caucasus, particularly Azerbaijan, are consistently carried out.
 
In particular, these studies encompass the geology and geodynamics of the Caucasus \cite{ISMAILZADEH2020, Martin_Bochud}, magmatism and heat flow, active tectonics, and tectonic stresses (e.g., \cite{ISMAILZADEH2020, Tsereteli_faults, Martin_Bochud, opentech}). They also address seismicity, gravity, and density models. Special attention is given to the exploration of active faults \cite{Faridi}, including the Kur Fault, a major thrust structure within the Kura fold-and-thrust belt in Azerbaijan \cite{Tibaldi2024}.
Notably, this region is not only seismically active but also home to critical infrastructure, such as the large Mingachevir and Shamkir hydroelectric water reservoirs and plants (e.g., \cite{Tibaldi2024, ICSF24paper, ActaGeo2025}).

In addition to geological, seismological, and other applied research, theoretical studies of earthquake sequences utilizing statistical methods have been developed. The statistical analysis of Azerbaijan's seismicity, observed from 2003 to 2016, was conducted by \cite{Telesca2017}. This analysis involved evaluating the frequency-magnitude distribution of earthquakes and fitting it to the Gutenberg-Richter law, constructing global and local coefficients of variation, extracting scaling exponents from magnitude time series, assessing the correlation dimension for the spatial distribution of epicenters, and examining other related parameters.
However, the phenomenon of earthquake clustering remains insufficiently explored, despite its identification during analyses of the spatial distribution of earthquake epicenters (e.g., \cite{Telesca2017, Tsereteli_faults, opentech}).

It should be noted that recent advances in cluster analysis \cite{Ansari_review} enable more effective studies of this issue by mitigating the impact of large uncertainties in data, non-quantitative and subjective analyses, and the misinterpretation of significant amounts of inaccurate information \cite{Zamani, Georgoulas}.

Numerous studies have been dedicated to earthquake clustering (e.g., \cite{Cesca_cluster, Yeck_cluster, Fana_cluster, Piegari_cluster, Taroni, Ansari2009, Vijay, Nicolis, Vallianatos, SCITOVSKI_cluster, Perezan2018, Perezan2024}). This highlights the immense potential of cluster analysis and machine learning techniques across various scientific disciplines, particularly in seismology.

This research is based on the studies outlined in \cite{geophys2025, ActaGeo2025}. It focuses on earthquakes that occurred in Azerbaijan and adjacent areas from 2010 to 2023, using an original procedure to cluster earthquake catalogs.
Initially, the presence of formal clusters within an earthquake sequence is roughly estimated using the Morisita Index (MI), which serves as an indicator of grouped data. Following this, the catalog is divided into clusters using the DBSCAN \cite{DBSCAN, SCITOVSKI_cluster} and HDBSCAN \cite{HDBSCAN} algorithms, with the results monitored through the Silhouette Index (SI).
Partitions providing the maximal SI or values close to it are then compared for similarity using the Adjusted Rand Index (ARI) \cite{Hubert, Moulavi2014, Channamma, DudekSlh}, which helps analyze and reduce the number of partitions by eliminating those with high similarity. Additionally, the resulting partitions are compared with the regional fault network.
This procedure enables a reliable identification of clusters representing earthquake epicenters or foci. Furthermore, the depth distributions of earthquake clusters are examined, and these analyses contribute to the development of a systematic understanding of seismic activity.

\section{The earthquake data selection}

We examine earthquakes occurring within the rectangular region bounded by latitudes 38$^\circ$ to 42$^\circ$ N and longitudes 45$^\circ$ to 50$^\circ$ E. From the catalog \cite{earthquakes_bulletin}, a total of 6194 earthquakes with magnitudes $M \geq 2$ and depths exceeding 0.1 km were identified. A comparable dataset of Azerbaijani earthquakes has also been analyzed by \cite{Telesca2017}. 
Each seismic event is defined by its latitude, longitude, depth, time, and various magnitude types.

This study does not involve filtering the catalog data. 
We are interested in the spatial clustering of hypocenters and their depth distribution, which typically lie beyond the reach of regular human activity. Therefore, it is important to have access to the broadest possible set of earthquake hypocenters. Moreover, the catalog data for this region is quite uneven across years, sources, and methods of earthquake identification, and not sufficiently comprehensive.
Thus, 
%As a rule, information extracted from earthquake catalogs requires 
catalog pre-processing \cite{Perezan2018}, which typically involves the removal of non-tectonic events \cite{Perezan2024, Mitchinson2024} and those assigned characteristics by the catalog operator \cite{Mitchinson2024}, is not performed.

An essential step in catalog processing is the verification of its completeness, which includes the estimation of the completeness magnitude $M_c$ \cite{Telesca2017, Perezan2018},
%
%This study does not involve filtering the catalog data. Instead, we directly proceed to estimate the completeness magnitude $M_c$,
 which represents the minimum magnitude above which all earthquakes are reliably recorded in the catalog. To determine $M_c$, we applied the MAXC method \cite{M0eval, Telesca2017}. Based on our previous analysis \cite{ActaGeo2025}, a completeness magnitude of $M_c = 2$ is appropriate, allowing us to include all 6194 earthquakes with $M \geq M_c$ in the analysis.

\begin{figure}[tbh]
\centering
\includegraphics[width=6 cm, height=4.5 cm]{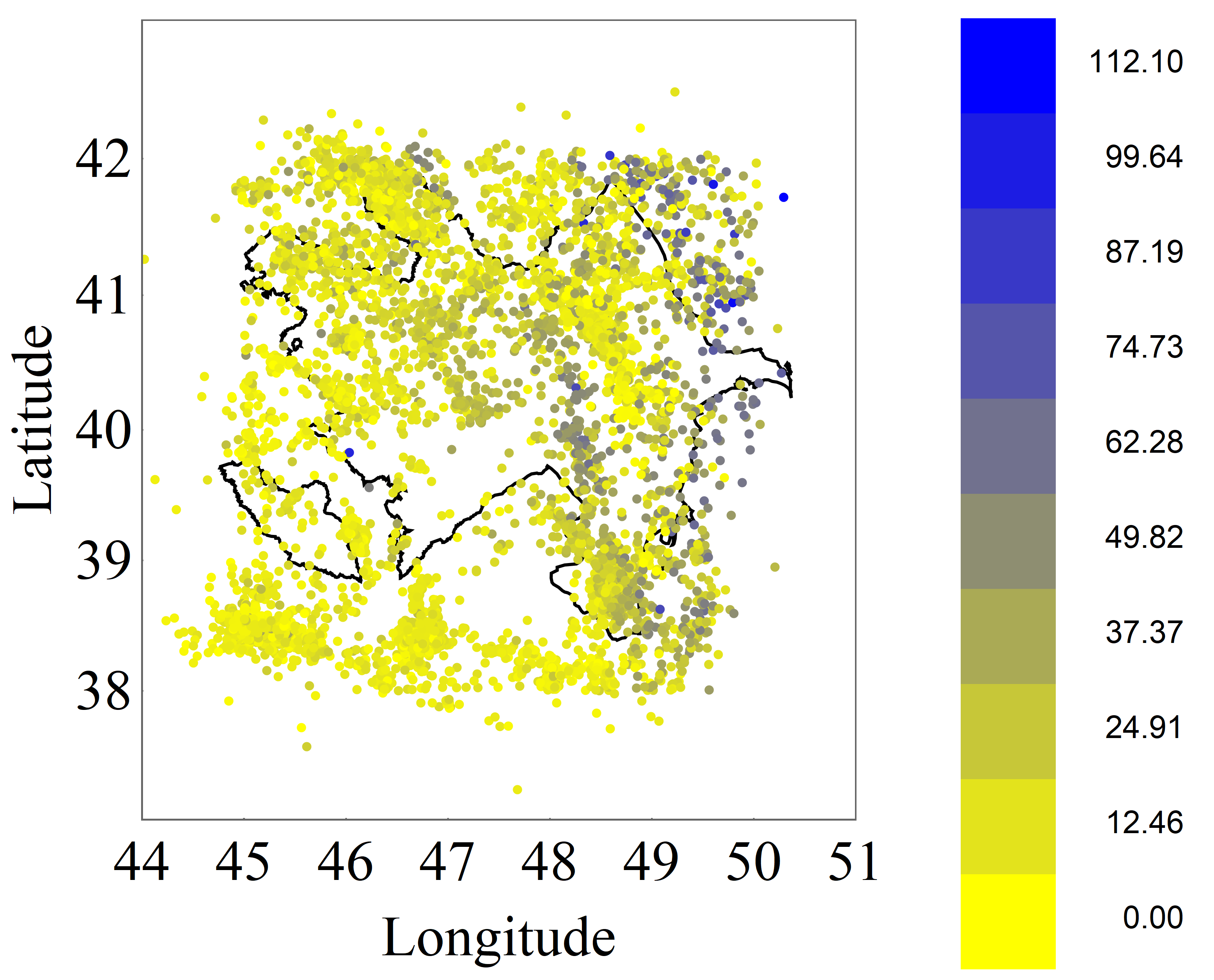}\hspace{0.2 cm}
\includegraphics[width=6 cm, height=4.4 cm]{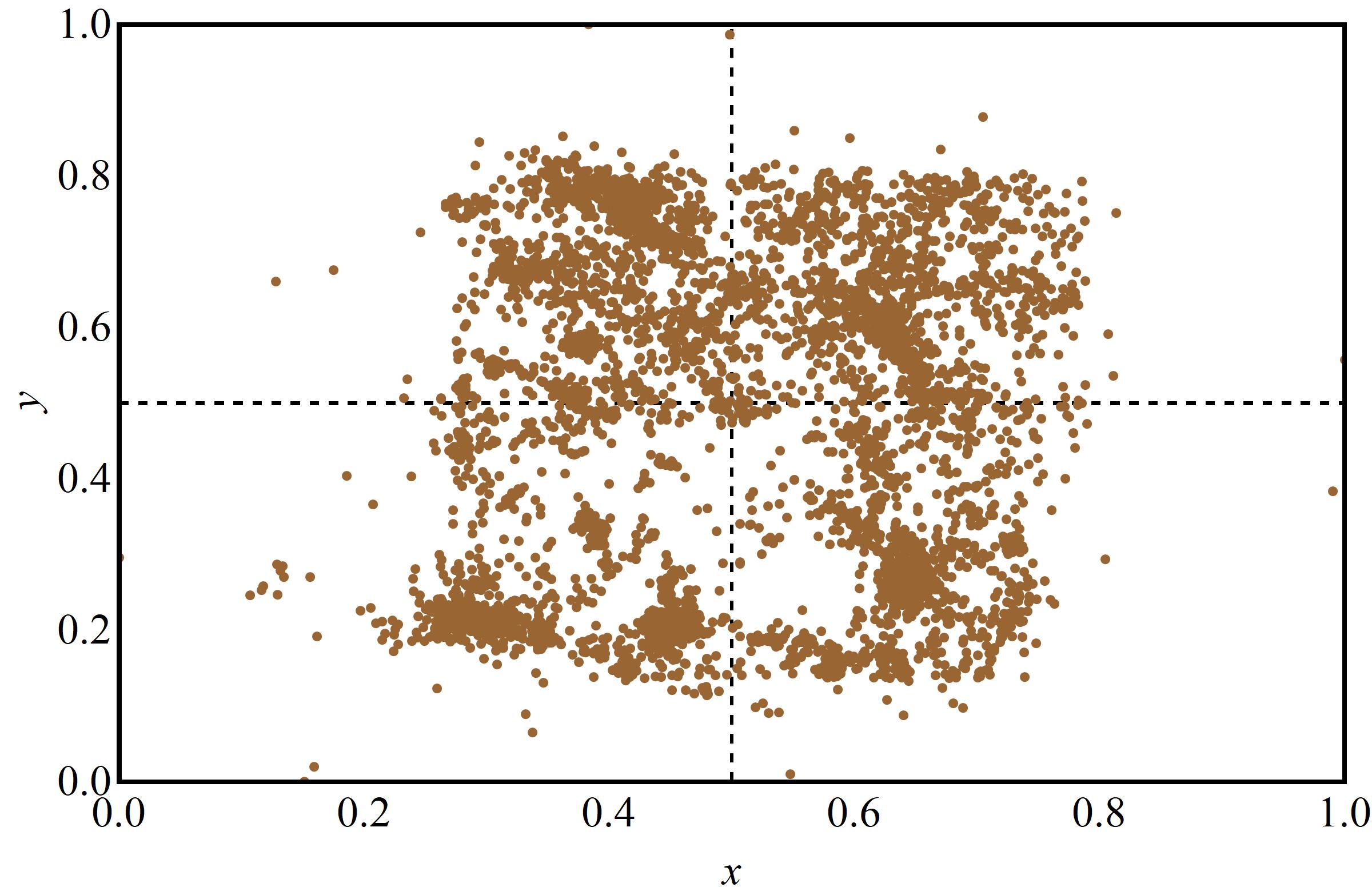} \\
(a) \hspace{6 cm}  (b)
\caption{a: Distribution of earthquake epicenters across Azerbaijan and surrounding regions. Earthquake depth is represented by color according to the accompanying colorbar. b:  Normalized locations of the earthquake epicenters corresponding to panel  (b).}
\label{fig:1}
\end{figure}

To provide an overview of the spatial distribution of the selected earthquakes, we plot the earthquake epicenters in Fig.~\ref{fig:1}a as filled circles, with color indicating the depth of each event. Most earthquakes are relatively shallow, while deeper seismic events are predominantly concentrated near the coast.

\section{Morisita index application}

Since our focus is on identifying earthquake groupings, it is appropriate to begin with a coarse criterion for clustering. To this end, we apply an approach that estimates the dispersion of spatially distributed points. Specifically, we use the box-counting method, in which the study area is divided into small cells, typically cubic  grids, and the number of points within each cell is counted. This provides a basis for introducing  a quantitative measure of spatial clustering in the form of the Morisita Index $I_\delta$, which, retaining the notation of works \cite{ouchi86, Telesca, Sharma}, is calculated as follows: 
\begin{equation}\label{skur:Morisita}
I_\delta=Q\frac{\sum_{i=1}^Q n_i (n_i-1)}{N(N-1)},
\end{equation}
where $\delta$ is the characteristic scale (side, diagonal, etc.) of a cell, $Q$ is the number of grid cells (boxes) covering the study area, $n_i$ denotes the number of points within the $i$th cell, and $N$ is the total number of points in the area. As mentioned in  \cite{ouchi86}, $I_\delta=1$  when points are  located completely randomly,  $I_\delta<1$ is observed for uniformly or regularly dispersed points, and finally,  $I_\delta>1$ means aggregated (clumped or clustered) point locations.

MI can be formally defined to multi-dimensional data. However, the computational complexity increases significantly with dimensionality. Therefore, it is practical to perform preliminary calculations using two-dimensional data representing earthquake epicenter locations, and subsequently compare the results with those obtained from three-dimensional data (hypocenter coordinates), where applicable.      

To simplify the evaluation of $I_\delta$, we transform the coordinates of earthquake hypocenters $(P, R, H)$ into the unit cube using the following mapping:
\begin{equation}\label{skur:trans}
x=\frac{P-P_{\min}}{ P_{\max} - P_{\min}},\quad 
y=\frac{R- R_{\min}}{ R_{\max}-R_{\min}}, \quad
z=\frac{H-H_{\min}}{H_{\max}- H_{\min}},
\end{equation}
where $\{P_{\min},P_{\max}\}=\{42.3021, 52.1283\}$, $\{R_{\min},R_{\max}\}=\{37.1765, 43.2183\}$, $\{H_{\min}$, $ H_{\max}\} = \{0; 112.1\}$.
Then each side of the cube is divided into $n$ parts of the length $\delta = 1/n$. 

\begin{figure}[tbh]
\centering
\includegraphics[width=8 cm, height=5.5 cm]{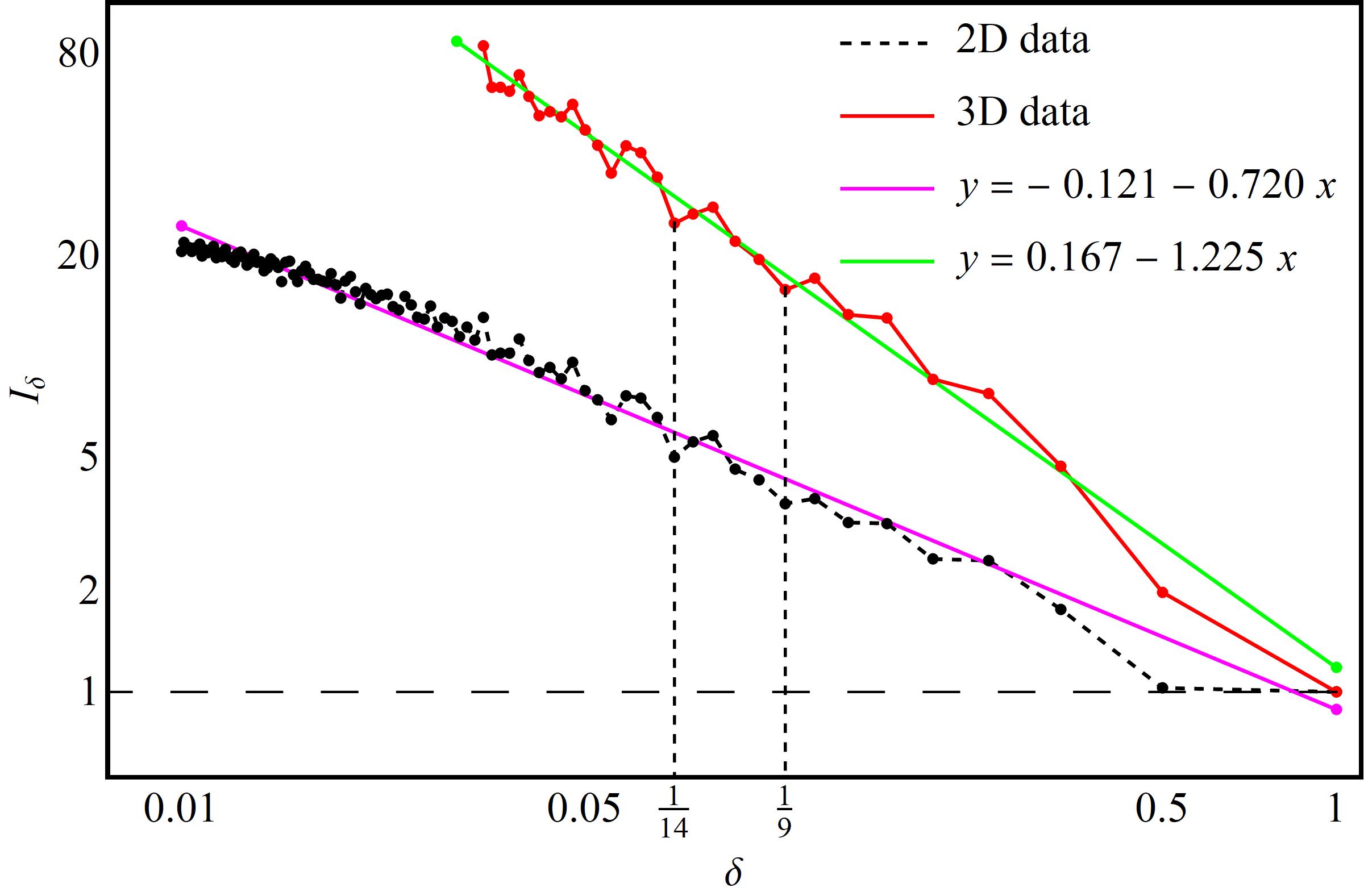}
%\\
%(a)\\
%\includegraphics[width=6 cm, height=4.5 cm]{figure2B_Morisita} \hspace{0.5 cm}
%\includegraphics[width=6 cm, height=4.5 cm]{figure2C_Morisita}\\
%(b) \hspace{6 cm}  (c)
\caption{Morisita Index $I_\delta$ vs. $\delta$, represented in a logarithmic scale and evaluated for 2D (black dashed) and 3D (red solid) data, respectively.  }\label{fig:2}
\end{figure}

We begin the MI calculation with the case of 2D data, which, after the transformation (\ref{skur:trans}), is distributed over the unit square, as shown in Fig.\ref{fig:1}b. The unit square is subdivided into $ n^2$ quadrats. The resulting Morisita index plot is presented in Fig. \ref{fig:2} (the lower dashed curve), illustrating the trend as the mesh size $\delta$ decreases from 1 to 1/100.  The graph is bounded on the right by the value $I_\delta=1$ at $\delta=1$.

%Transforming the locations of earthquake hypocenters using relation (\ref{skur:trans}),
Similarly,  we perform  MI calculations for the 3D data, which are substantially more computationally intensive, and plot the corresponding upper  broken line in Fig.~\ref{fig:2}. %Unfortunately, the calculation of $I_\delta$ was terminated after $n=30$ due to the high computational demands. Nevertheless, 
Comparing the shapes of the $I_\delta$ curves obtained for the 3D and 2D data reveals a clear similarity.

Thus, since the MI exceeds 1, we can conclude that earthquakes are not distributed entirely randomly and tend to form clusters. Before proceeding to cluster identification, let us consider several features of MI.

First, it is interesting to clarify the nature of ``pits'', which are observed on the $I_\delta$ profiles (Fig.\ref{fig:2}), evaluated for both 2D and 3D data  at $\delta=1/9$ (poorly distinguishable), $\delta=1/14$ (more visible), etc.  Analyzing the mutual position of the mesh and the dataset, we noticed that for cell sizes $\delta=1/9$ or $\delta=1/14$  the mesh nodes fall inside some clusters, whereas for the larger values of $\delta$, i.e. 1/8 or 1/13, these clusters lie inside the mesh cell.

This can be explained qualitatively as follows.
Consider the partial cell formed by the less dense mesh, assuming that it contains $r$ cluster points and neglecting the points  outside this cell.  The denser mesh subdivides this cell into four smaller cells, each containing fewer cluster points, denoted as $r_{1,2,3,4}$, such that $r = \sum r_i$.   
  Then the expression (\ref{skur:Morisita}) for $I_{\delta=1/8}$ incorporates the term $r(r-1)$, while $I_{\delta=1/9}$ contains $ \sum r_i(r_i-1)$. Let us examine the difference between these terms and, by performing elementary transformations, prove that it is possitive, i.e.,   $r(r-1)-\sum r_i(r_i-1)=r^2-\sum r_i^2>0$. From this it follows that  $I_{\delta=1/8}>I_{\delta=1/9}$. 
  For 3D case, the mesh node falling into a cluster generates nine smaller mesh cubes. This reduces  the number of cluster points that fall into smaller mesh cubes   to a greater extent than in the 2D case.  Therefore, the deeper and more pronounced ``pits'' are observed for the 3D case.      
  The above considerations offer a qualitative framework for understanding how the Morisita Index effectively distinguishes a sufficiently contrasting cluster from a background of sparsely distributed points.

%\caption{a: Morisita Index $I_\delta$ vs. $\delta$. Here $m_1=8$ and $m_2=12$. Inset shows the sequence $L_n=\ln I_{1/n} /\ln n$, defined in (\ref{skur:lim}), approaching its constant value (green). The plots are  evaluated for 2D (black) and 3D (red) data, respectively.    b: The part of Fig.\ref{fig:1}b with meshes  corresponding to $\delta=1/m_2=1/12$ (red) and $\delta=1/11$ (green).}

%Transforming the locations of earthquake hypocenters using relation (\ref{skur:trans}), we perform similar MI calculations for the 3D data, which allows us to construct the upper broken line in Fig.~\ref{fig:2}a. Unfortunately, the calculation of $I_\delta$ was terminated after $n=30$ due to the high computational demands. Nevertheless, comparing the shapes of the $I_\delta$ curves obtained for the 3D and 2D data reveals a clear similarity. Moreover, the "pits" become more pronounced, although their positions remain unchanged.ronounced, while their positions remain unchanged.

The second MI feature that we want to discuss concerns the asymptotic index behavior.
MI analysis provides additional insight into the fractal nature of datasets \cite{Perezan2018}. The method of MI calculation, namely, the box-counting approach, is also commonly used to assess the dimensions of fractal objects. Therefore, it is not surprising that MI is related to the fractal dimensions of self-similar point patterns. As shown in \cite{Telesca}, the relationship between MI and the fractal dimension is established through an asymptotic power-law dependence, $I_\delta \propto \delta^{p}$ ($p<0$). According to \cite{Telesca}, the exponent $p$ is related to the second-order Morisita slope (also referred to as the codimension of order 2) by $p = -S_2$. Furthermore, the relation $D_2 - E = p$ holds, where $D_2$ is the second-order R\'{e}nyi generalized dimension (or correlation dimension) and $E$ is the dimension of the Euclidean space in which the points are embedded.

%In addition, the following relation is valid (with $\delta = 1/n$):
%\begin{equation}\label{skur:lim}
%\lim_{n\to \infty}  L_n=\lim_{n\to \infty}  \frac{\ln I_{1/n}}{\ln n}=p=const.
%\end{equation}
To verify this property, %(\ref{skur:lim}), 
we consider the asymptotic behaviour of the sequence  $I_{\delta}$ plotted in a logarithmic scale (Fig.\ref{fig:2}). To do this, the linear regression line $y=-0.121 - 0.720 x$
 evaluated by the least-squares method for the 2D data.  Corresponding 95\% confidence interval for the line slope is
$\{-0.743, -0.697\}$ and the coefficient of determination is $R^2=0.975$. Thus, $p \approx -  0.72$.

For the 3D data, the  linear regression is $y=0.167 - 1.225 x$, 95\% confidence interval  for the slope $\{-1.28912, -1.1612\}$, and $R^2=0.982$. Thus, $p \approx -  1.225$.

  % The resulting dependencies are shown in the inset of Fig.~\ref{fig:2}a. It can be observed that $L_n$ tends toward a constant value, with $p \approx -0.68$ for the 2D data and $p \approx -1.27$ for the 3D data. This indicates that the spatial distribution of earthquakes exhibits fractal properties.

It is worth noting that \cite{Telesca2017} calculated the correlation dimension for spatially distributed epicenters to be approximately $D_c \approx 1.38$. Our evaluations yield a close estimate of the fractal dimension, with $D_2 = 2 + p = 1.28$. For the 3D case, the corresponding dimension is $D_3 = 3 + p = 1.775$.   %which is not very typical for 3D fractal objects. This probably can be related to the not enough number of points in the dataset or its multifractal nature when the box-counting algorithm catches a partial fractal.   

Thus, MI calculations show that the earthquake dataset exhibits fractal properties and is characterized by a pronounced cluster structure, for which a spatial scale can be assessed. Further studies concern the extraction of spatial clusters from the dataset utilizing the corresponding algorithms.

\section{DBSCAN and HDBSCAN application}

To perform the cluster analysis, we use the well-known DBSCAN algorithm and its improved version, HDBSCAN, implemented via the {\it sklearn.cluster} module in Python \cite{scikit-learn}. Some preliminary results concerning the application of DBSCAN to this catalog were previously presented in \cite{geophys2025}. %Here, we summarize only the key points from \cite{geophys2025}.

According to the recommendations of \cite{Piegari_cluster}, the data prior to clustering should be scaled such that each component is transformed to a common range. We use transformation (\ref{skur:trans}), which provides the translation of each coordinate into the range $(0;1)$.

Next, the DBSCAN algorithm is applied to the scaled catalog. Recall that DBSCAN \cite{DBSCAN,Piegari_cluster,SCITOVSKI_cluster} requires two intrinsic parameters: $\varepsilon$ and 
{\it min\_samples}. The former defines the $\varepsilon$-neighborhood of each point, specifying the maximum distance between any two points within the neighborhood. The latter sets the minimum number of points required within the $\varepsilon$-neighborhood to form a cluster. To estimate appropriate ranges for these parameters, we apply the approach proposed in \cite{DBSCAN}, with an adaptation detailed in Appendix \ref{skur:appA}.  Note that both parameters influence the density and size of a cluster. Similarly, the Morishita index encapsulates information related to clustering and provides a rough estimate of cluster size. This parallel suggests a potential relationship between the DBSCAN parameters and the Morishita index.  However, this subtle issue remains beyond the scope of these studies.

Thus, the parameter $\varepsilon$ is selected from the range $\varepsilon \in [0.01, 0.2]$ with a step size of 0.01, and {\it min\_samples} is chosen from the range $[80, 200]$ with a step size of 10. The {\it euclidean} distance metric is used, which for two points $x(x_1, x_2, x_3)$ and $y(y_1, y_2, y_3)$ is calculated  as  $d(x,y)=\sqrt{\sum_{i}^3{(x_i-y_i)^2}}$. 
%resulting in a total of 242 cases.   

The result of DBSCAN is a partition consisting of clusters along with a set of unclassified points (noise points). It is important to note that different combinations of the parameters $\varepsilon$ and {\it min\_samples} can yield partitions with the same number of clusters.

The quality of point classification is assessed with the help of Silhouette
Index \cite{Slh}. Let us recall that  SI of  the set $C$, which is partitioned by $N $ clusters $C_i$, $i=1,\dots,N$,  is evaluated as follows 
$\text{SI}(C)=\sum_{i=1}^{N}s(i)/N$, where 
 \begin{equation*}
 s(i)=\frac{a(i)-b(i)}{\max\{a(i),b(i)\}}.
 \end{equation*}
 Similarly,  the Silhouette Index  is evaluated for each cluster $C_i$ according to the relation  
 \begin{equation*}
 \text{SI}(C_i)=\frac{1}{|C_i|}\sum_{j=1}^{|C_i|} s(j),
 \end{equation*}
 where $|C_i|$ is the number of elements in the $C_i$ cluster.
 The
 mean intra-cluster distance $a(i)$ and mean nearest-cluster distance $b(i)$ are calculated via the following relations
 \begin{equation*}
  a(i)=\frac{1}{|C_i|-1}\sum_{j=1}^{|C_i|} d(i,j), \qquad b(i)=\min_{k\neq i} \sum_{j=1}^{|C_k|} d(i,j),
 \end{equation*}
 where $d(i,j)$ is a distance.

 When all clusters are well separated and dense enough, SI($C$)   is close to 1. The low positive $\text{SI}(C)$ value indicates poorly separated clusters. When $-1<\text{SI}(C)<0$, the elements refer to misclassified. 
 In fact,  SI technique provides  good estimation of clustering quality when clusters are well localized and convex, and  they should be linearly separable. SI doesn't capture well the clusters of complex shapes or varying densities. In particular, such a situation occurs when DBSCAN algorithms are used. Then, to evaluate the clustering results, we employed various approaches and alternative indices \cite{Arbelaitz_index_rev,Scitovskibook21}, including the evaluation of similarity between two partitions using the Adjusted Rand Index (ARI), which we apply below.

\begin{figure}[tbh]
\centering
\includegraphics[width=6 cm, height=4.5 cm]{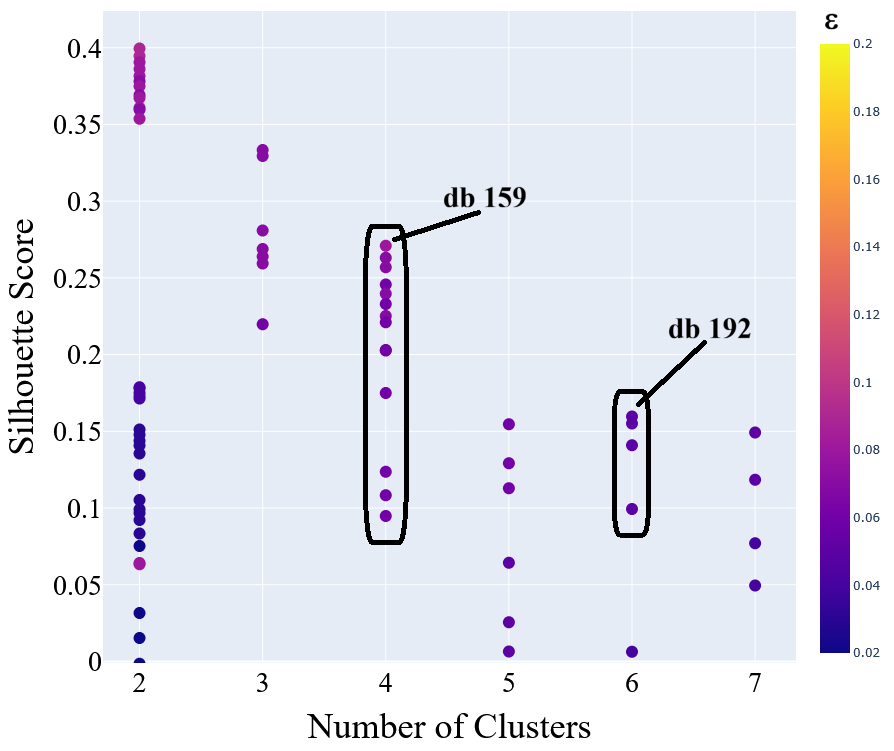}\hspace{0.2 cm}%figure3_afigure3_b
\includegraphics[width=6 cm, height=4.5 cm]{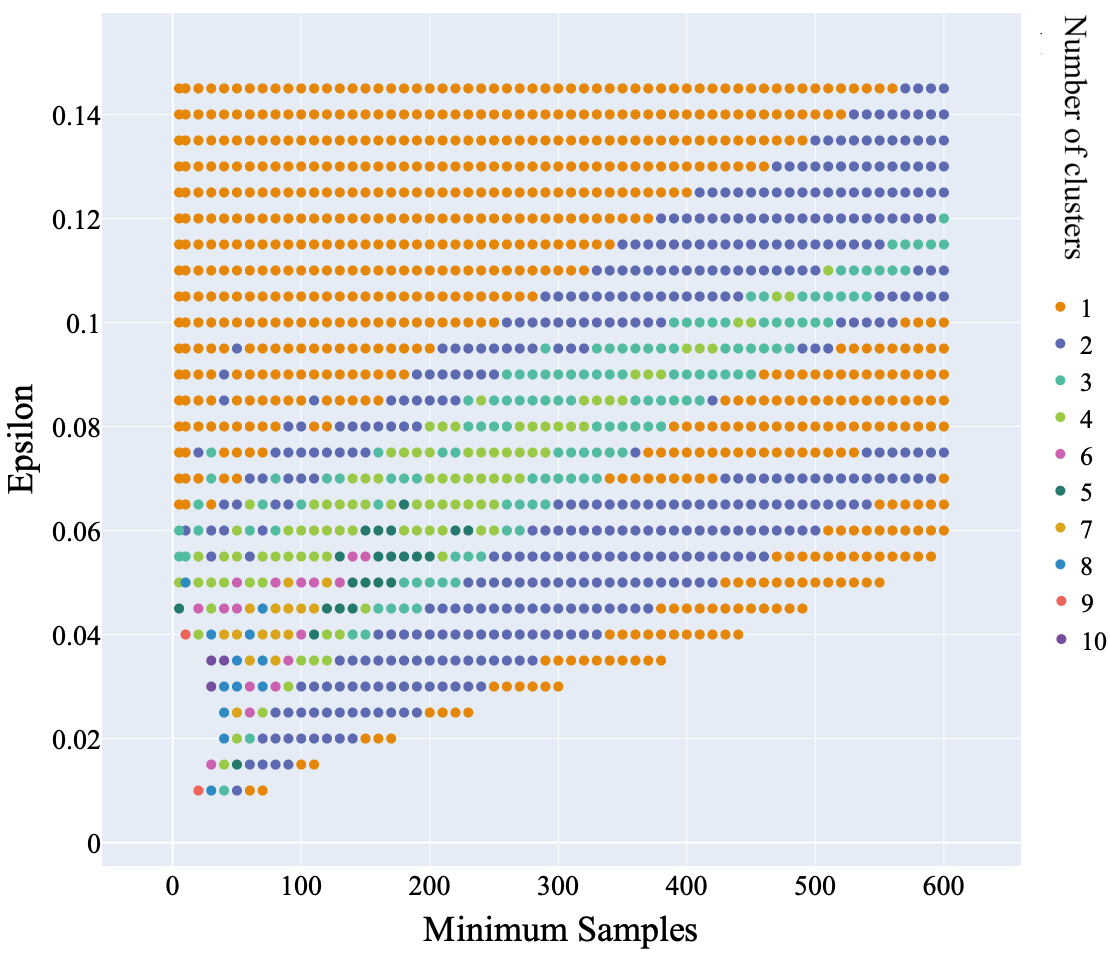} \\
(a) \hspace{6 cm}  (b)\\
\includegraphics[width=6 cm, height=4.5 cm]{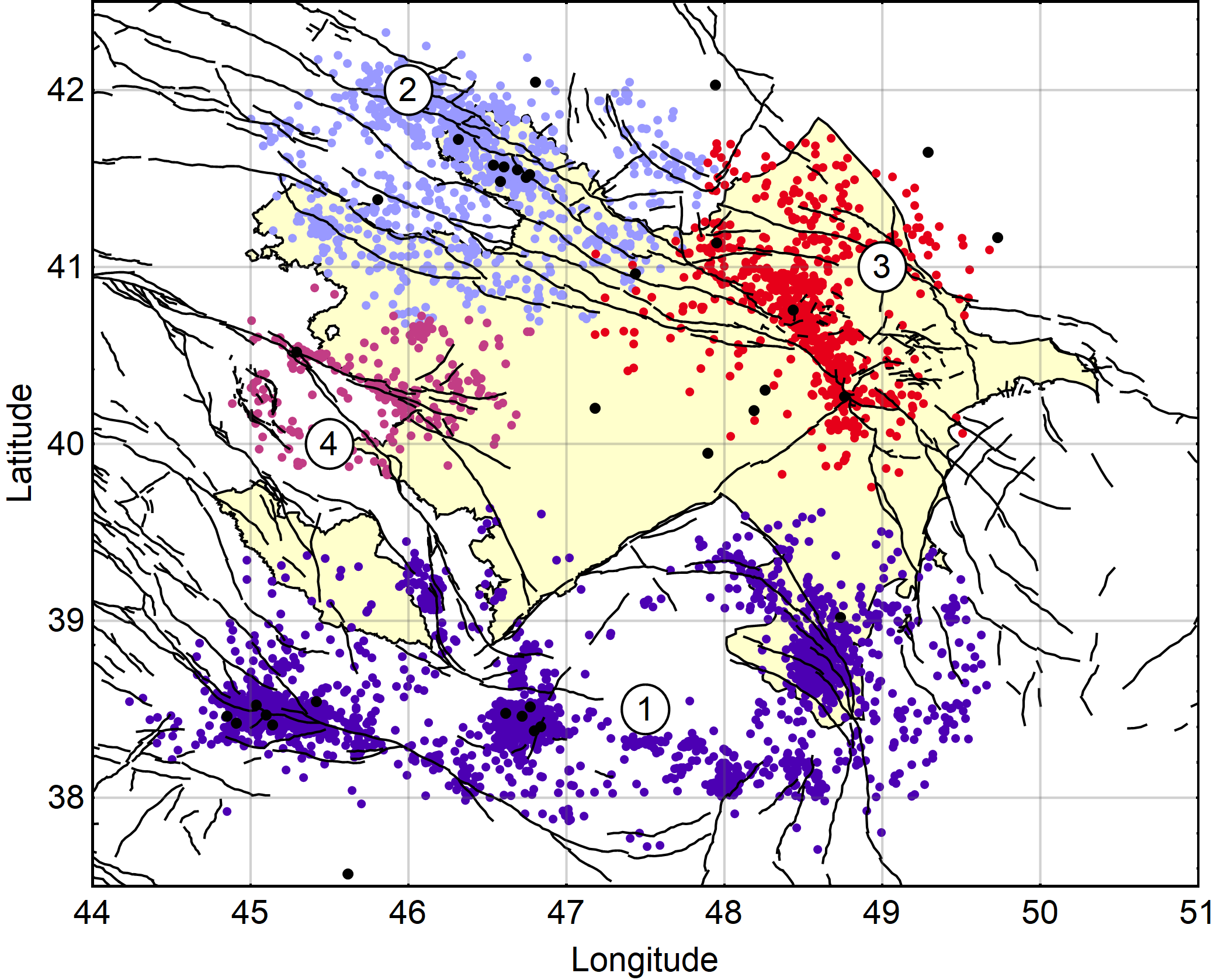} 
\includegraphics[width=6 cm, height=4.5 cm]{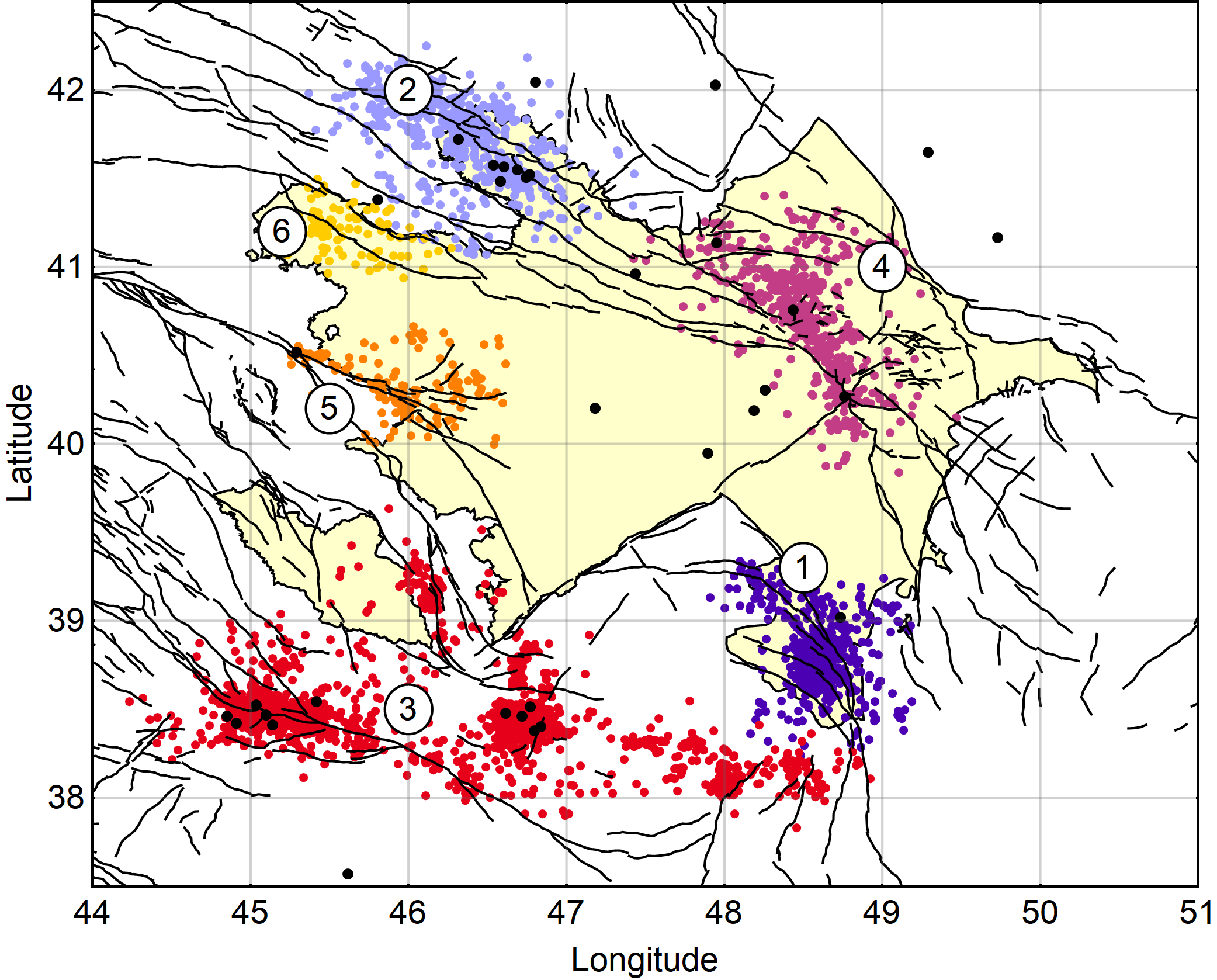} \\
(c) \hspace{6 cm}  (d)\\
\caption{a: Silhouette Index  vs. number of clusters.  b: Two-parameter diagram showing  the distribution of  cluster counts. c and d: The 4- and 6-cluster partition marked in the Silhouette diagram by the labels ``db 159'' and ``db 192'', respectively. The noise points are removed. Filled bullets mark strong earthquakes with $M>5.5$. The results obtained via DBSCAN. }\label{fig:3}
\end{figure}

Thus, the calculation of SI for each partition results in the diagram  shown in Fig.\ref{fig:3}a. To consider the influence of
the parameter $\varepsilon$ to clustering, we associate point positions with a colorbar
such that the lighter points correspond to higher values of $\varepsilon$. From Fig.~\ref{fig:3}a it follows that SI decreases with increasing number of clusters in partitions.  The parameters of partitions provided SI$_{\max}$ are included in Table \ref{skur:Table01}.

It is useful to find out the effect of both parameters $\varepsilon$ and  {\it min\_samples} on the number of clusters generated by DBSCAN. To do this,  the diagram (Fig.~\ref{fig:3}b) representing the  numbers of clusters in the partitions depending on  $\varepsilon$  and {\it min\_samples} is constructed on a dense grid,  where $\varepsilon$ varies in the range $[0.01;0.15]$ with a step of 0.005 and   {\it min\_samples} ranges from $[10;600]$ with a step of 10.
This shows that there is a bounded  region in the two-parameter space whose points correspond to cases where there is more than one cluster in the partition. It has a complex structure, especially for small values of $\varepsilon$ and  {\it min\_samples}, and degenerates into a narrow band  as the parameters increase. 

Let us consider in more detail, the partitions with a medium number of clusters, e.g., cluster 4 (label ``db 159'') and cluster 6 (label ``db 192'') enclosed in  rectangles in Fig.\ref{fig:3}a. 
The locations of the clusters are depicted in Figs.\ref{fig:3}c and d.

\begin{table}[tbh]
\caption{The DBSCAN's parameters corresponding to the maximum SIs depicted in Fig.\ref{fig:3}a. }\label{skur:Table01}
\centering 
\begin{tabular}{@{}lllllll@{}}%{ c|c|c|c|c|c } 
 \toprule
 Number of clusters& 2 & 3 & 4 		& 5 		& 6 			& 7 		  \\ \midrule
 $\varepsilon$ &    0.09 &0.07&      0.08 	& 0.06 	& 0.05 	& 0.05 	 \\  \midrule
 {\it min\_samples} & 190& 120 & 200 	& 150 & 80 	& 90 	\\ 
 \bottomrule
\end{tabular}
\end{table}

\begin{figure}[tbh]
\centering
\includegraphics[width=6 cm, height=4.5 cm]{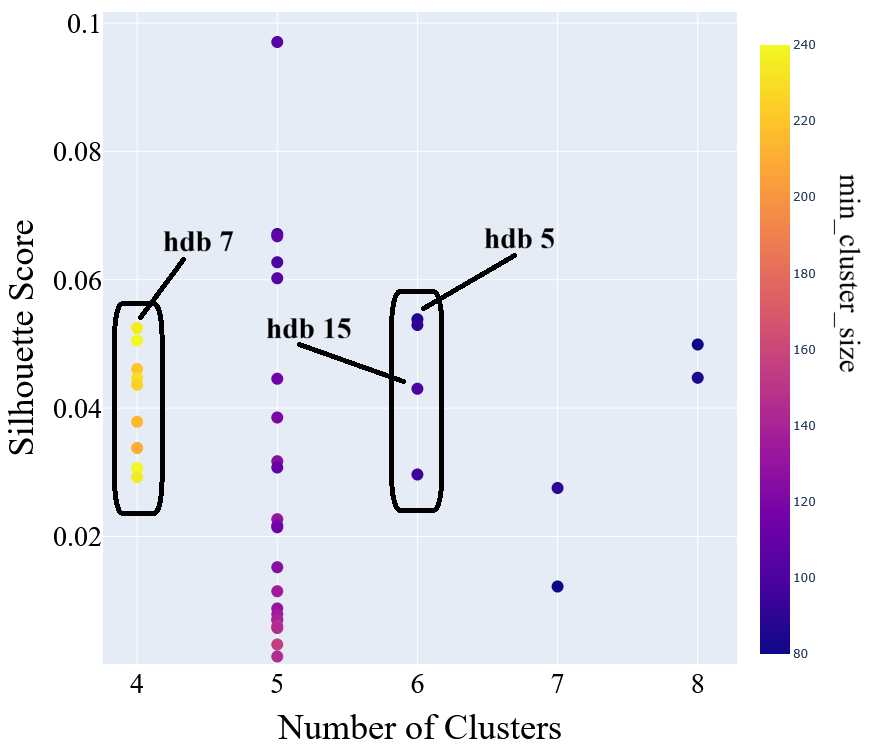}\hspace{0.2 cm}%
\includegraphics[width=6 cm, height=4.5 cm]{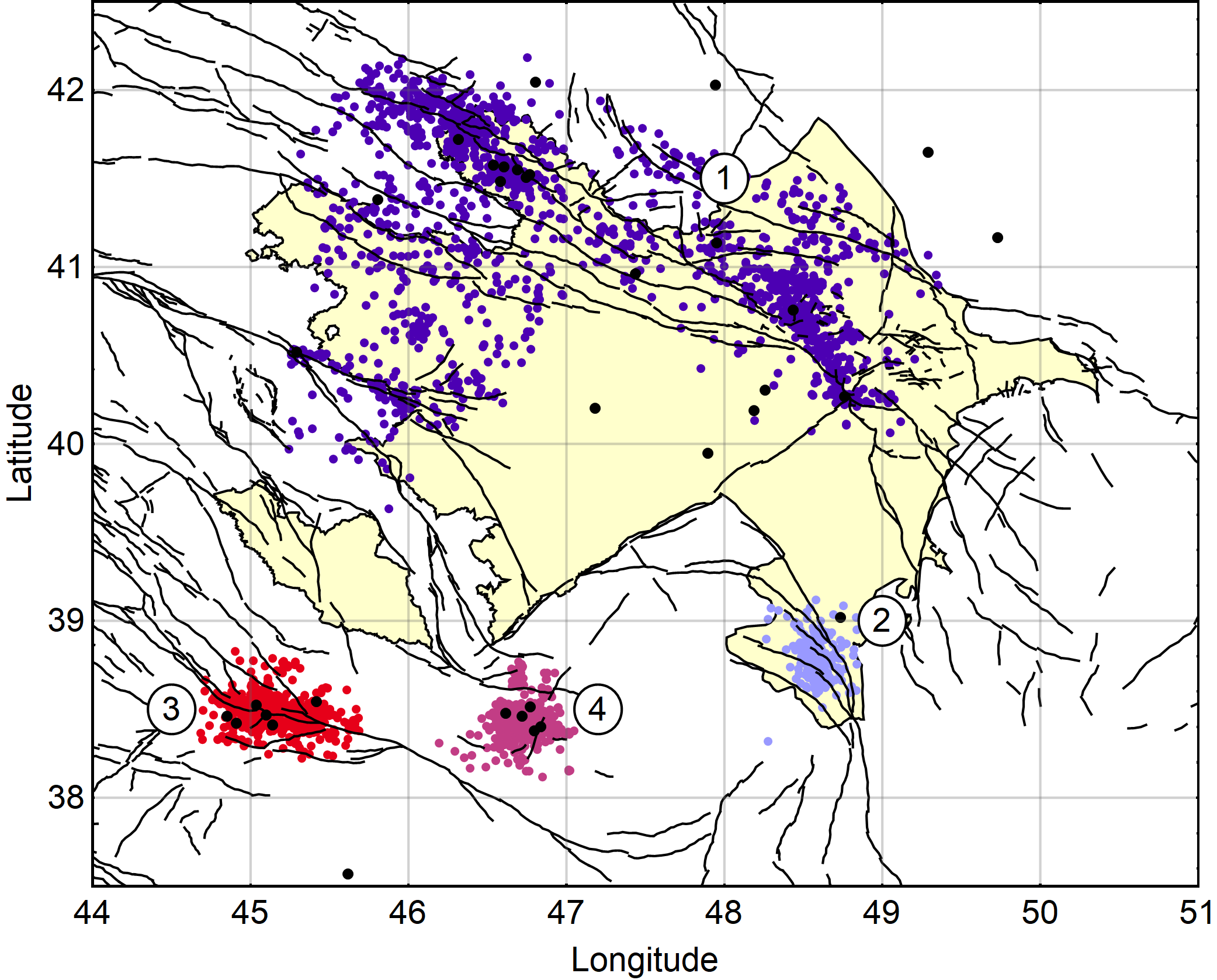} \\
(a) \hspace{6 cm}  (b)\\
\includegraphics[width=6 cm, height=4.5 cm]{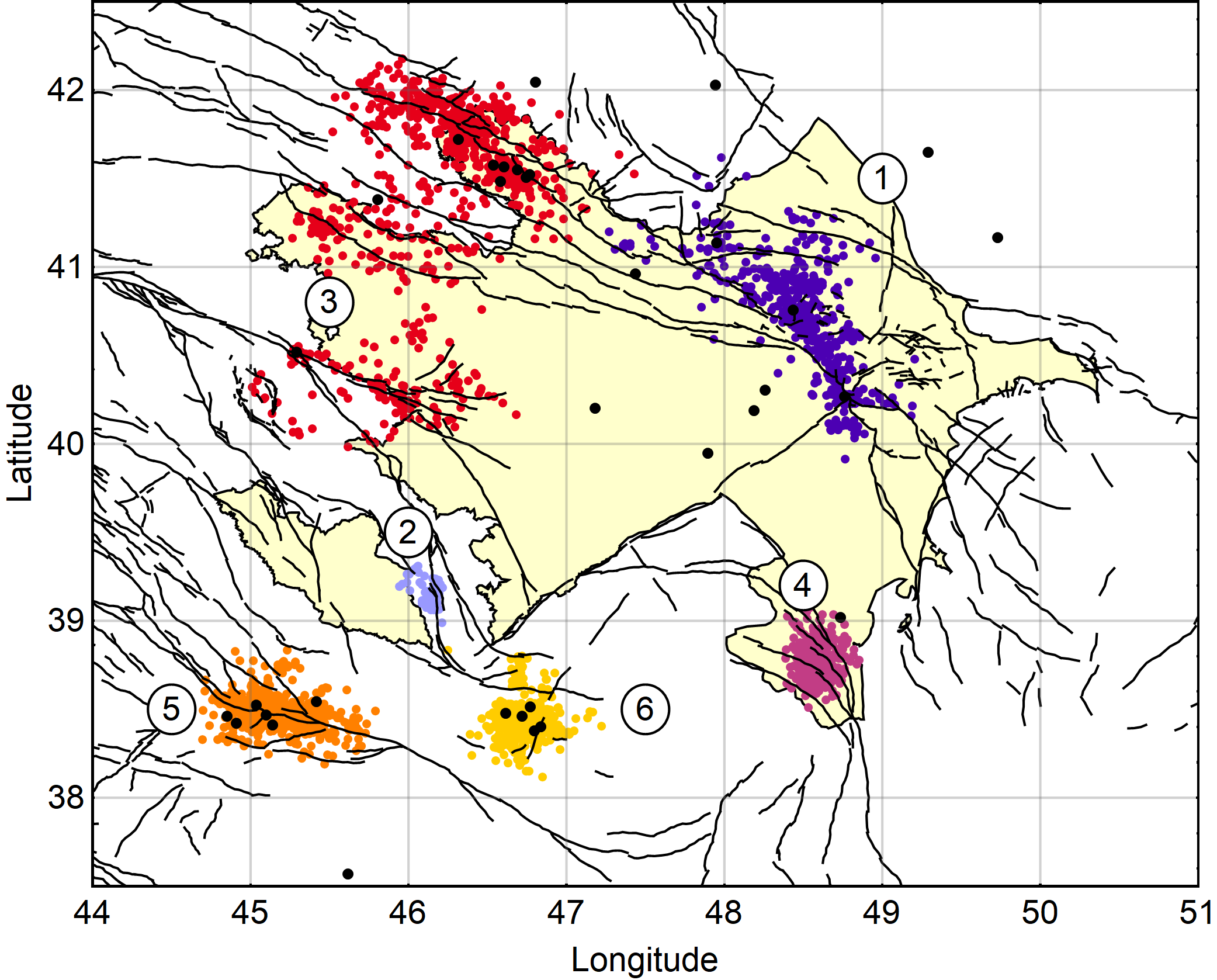} \hspace{0.2 cm}
\includegraphics[width=6 cm, height=4.5 cm]{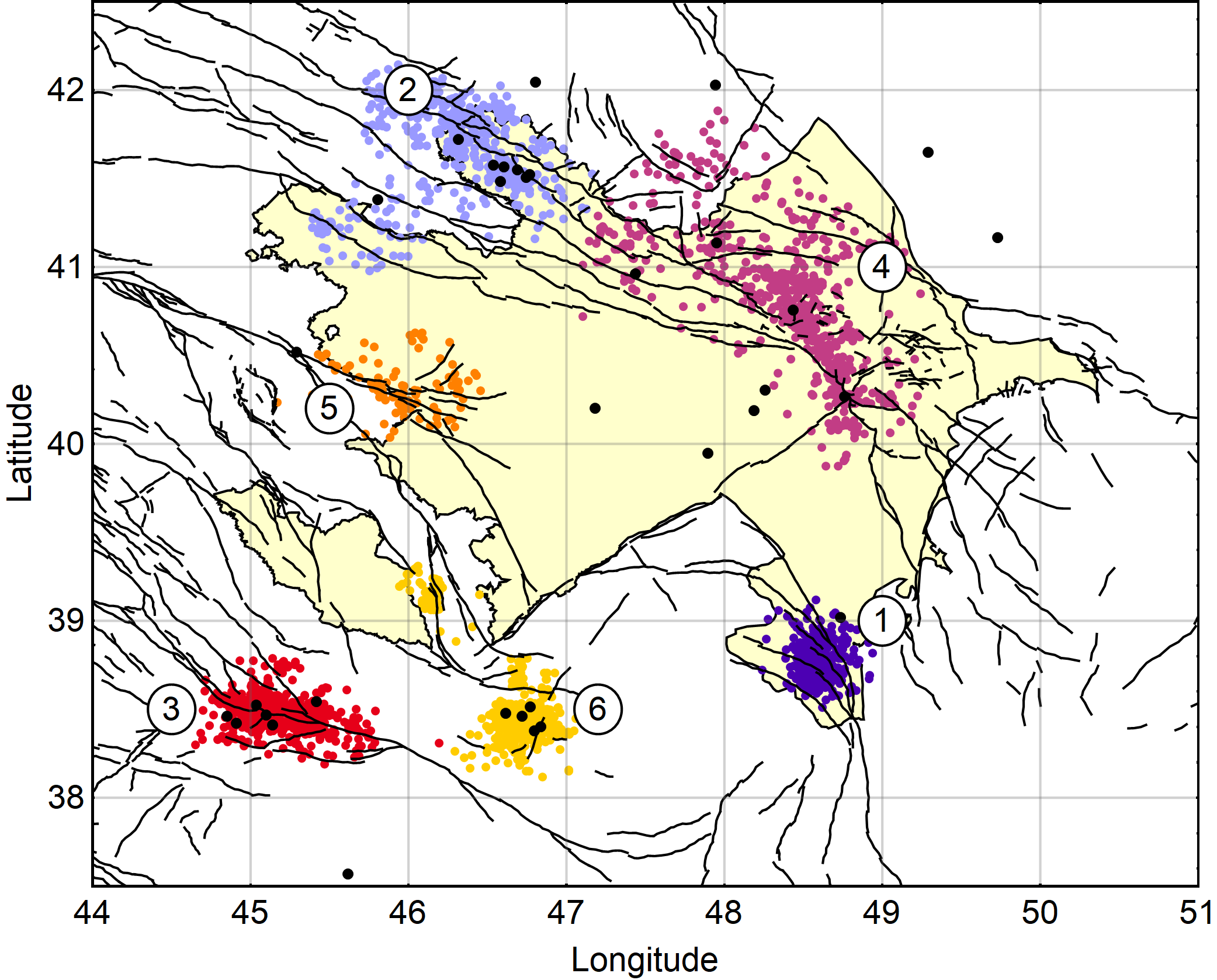} \\
(c) \hspace{6 cm}  (d)\\
\caption{a: Silhouette Index  vs. number of clusters. b, c, and d: The 4- and 6-cluster partitions with  noise points removed, marked in the Silhouette diagram by labels ``hdb 7'', ``hdb 5'', and ``hdb 15'', respectively. Filled bullets mark strong  earthquakes with $M>5.5$. The results obtained via HDBSCAN.  }\label{fig:4}
\end{figure}

Given that earthquake catalogs are often noisy and contain outliers, analyzing them with multiple algorithms is beneficial. For this reason, we also employ  the HDBSCAN algorithm. It can work with  clusters of varying densities, whereas DBSCAN tends to perform better with clusters of roughly uniform density. On the other hand, HDBSCAN may select clusters that are too dense and contain  a small number of elements, that, in turn,  reduces the possibility of conducting statistical analysis. Accordingly,  we apply both methods and compare their  clustering results.

Thus, let us perform similar evaluations of partitions using the HDBSCAN algorithm. We select the intrinsic parameter {\it min\_cluster\_size} from the range $(80, 240)$ with a step of 5, while the parameter {\it min\_samples} is set to its default value (equal to {\it min\_cluster\_size}), and either the  \textit{euclidean} or \textit{manhattan} metric (i.e., 
$d(x(x_1, x_2, x_3), y(y_1, y_2, y_3))=\sum_{i}^3{|x_i-y_i|}$) is used. %The total number of partitions generated is {\bf !!!!122}. 
 The SI values of  each partition are calculated  and the resulting points are shown in Fig.\ref{fig:4}a. In the partitions with the same number of clusters, SI$_{\max}$ is achieved.   For the specified parameter {\it min\_cluster\_size} range, the largest SI$_{\max}$ is observed for 5-cluster partition, in contrast to the DBSCAN results (Fig.\ref{fig:3}a).  For comparison with DBSCAN results, we  consider the 4-cluster partition  marked with the label of ``hdb 7'' (Fig.\ref{fig:4}a). The corresponding cluster location is plotted in Fig.~\ref{fig:4}b. We also consider two 6-cluster partitions labeled  ``hdb 5'' and ``hdb 15'' in the diagram of Fig.\ref{fig:4}a, which possess nearly identical SIs but differ in the composition of several clusters (Fig.~\ref{fig:4}c and d).

\section{Adjusted Rand Index analysis}

To quantify the differences between the partitions, we use the Adjusted Rand Index (ARI) \cite{Scitovskibook21}, which measures the similarity between two clusterings. Generally, these two partitions may consist of different numbers of subsets; however, in this study, we compare partitions containing the same number of clusters.
It is important to note that the ARI ranges from $-1$ (indicating particularly discordant partitions) to 1 (indicating identical partitions). For the ARI evaluation, we use the {\it adjusted\_rand\_score} function from the {\it sklearn.metrics} module \cite{ARI}. This approach helps us reduce the number of partitions that need to be considered as different.

\begin{figure}[tbh]
\centering
\includegraphics[width=12 cm, height=8 cm]{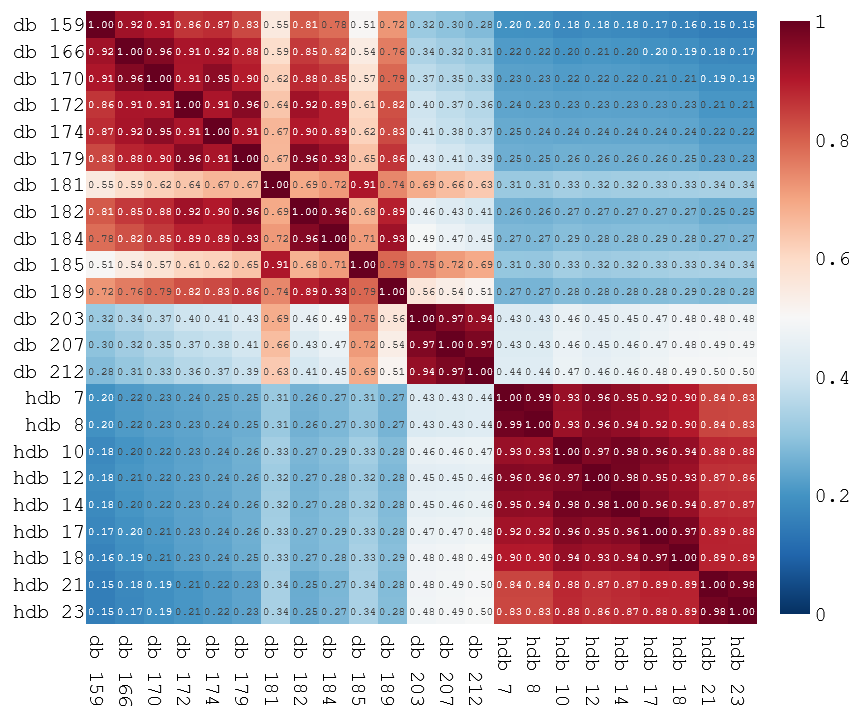}\\
(a)\\
\includegraphics[width=8 cm, height=5 cm]{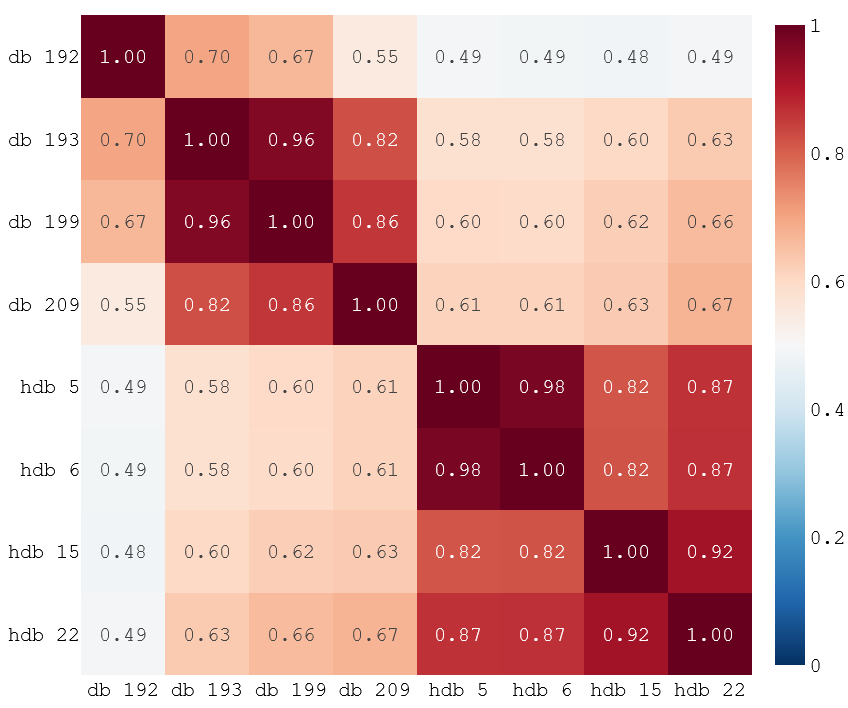} \\
%(a) \hspace{5 cm} 
 (b)
\caption{Adjusted Rand Index evaluation for 4-cluster  (a) and 6-cluster partitions (b). The axis labels correspond to the selected partitions in Figs.~\ref{fig:3}a and \ref{fig:4}a, numbered in descending order of SI.  The labels with ``db'' and ``hdb''  stand for  the results obtained via DBSCAN and HDBSCAN, respectively. }\label{fig:5}
\end{figure}

In the following, we select partitions with a fixed number of clusters, $N$, which are compactly located in the  rectangles depicted in the SI diagrams. 
We consider partitions with $N=4$ and $N=6$ clusters, selecting the subset enclosed in the rectangles shown in Fig.\ref{fig:3}a and Fig.\ref{fig:4}a. 
The results of the ARI evaluation for pairs of clusterings obtained for $N=4$ and $N=6$ are presented in matrix form in Figs.\ref{fig:5}a and \ref{fig:5}b, respectively. The row and column labels correspond to the partitions, for which the ARI is evaluated, listed in the order of their appearance within the specified rectangles in Fig.\ref{fig:3}a and Fig.\ref{fig:4}a. Labels ``db '' and ``hdb'' stand for partitions evaluated by DBScan and HDBScan method, respectively.  The evaluated ARI values are placed in the corresponding cells and highlighted by color. ARIs for identical partitions are equal to 1 and are depicted in dark colors, while other combinations yield smaller ARIs, represented in lighter shades.
  
Analyzing Fig.\ref{fig:5}a, we see that 14 four-cluster partitions (labels ``db 159 '', $\dots$, ``bd 212'') involve the set of similar partitions (labels ``db 159'' -- ``db 179'',  ``db 182'' , ``db 184'', and ``db 189'') with ARI varying within the range $[0.72;1]$. However, there is another set of partitions, i.e. labels ``db 181'' and ``db 185'', which are similar to each other with ARI=0.91, while similarity to other partitions varies in the range $[0.51;0.79]$. The group of partitions ``db 203''-- ``db 212'' consists of very similar partitions with ARI equal to 0.94 and 0.97, while they   are similar to other partitions with the ARI range $[0.28;0.56]$ that means, in fact, weak similarity. Nine partitions produced by HDBSCAN (labels ``hdb 7'', $\dots$, ``hdb 23'') are also very similar to each other with ARI $\in [0.83;1]$.
The comparison of pairs of partitions belonging to the sets of partitions  obtained by the different algorithms shows that they differ significantly. For such cases,   ARI varies from 0.49 to 0.15.

Interestingly, when we consider the 6-cluster partitions (Fig.\ref{fig:5}b), ARI indices are quite high both for the partitions obtained by the same algorithm and   for the partitions produced by different ones. This can mean, in particular, that the algorithms we applied provide increasingly similar partitions as the number of clusters in them increases.

%  We observe that the ARI varies within the range $[0.86;1]$ for the diagram in Fig.\ref{fig:5}a, within $[0.92;1]$ for the diagram in Fig.\ref{fig:5}b, and within $[0.72;1]$ for the diagram in Fig.~\ref{fig:5}c. 

%\begin{figure}[tbh]
%\centering
%\includegraphics[width=6 cm, height=3.5 cm]{ARI_dbs_4}\hspace{0.2 cm}%figure5_a
%\includegraphics[width=6 cm, height=3.5 cm]{ARI_dbs_6}\\
% (a) \hspace{5 cm}  (b)\\
%\includegraphics[width=6.5 cm, height=3.5 cm]{figure5_c} \\
 %(a) \hspace{3.5 cm}  (b) \hspace{3.5 cm} 
 %(c)
%\caption{Adjusted Rand Index evaluation for 4- (a), 6- (b), and 7-cluster partition (c). The axis labels correspond to the selected partitions in Fig.~\ref{fig:3}a, numbered in descending order of SI, in particuler, the partition marked by ``$\times$''  correponds to  the label 12 in the panel (b). The results obtained via DBSCAN. }\label{fig:5}
%\end{figure}

%\begin{figure}[tbh]
%\centering
%\includegraphics[width=6 cm, height=3.5 cm]{ARI_hdb_4}\hspace{0.2 cm}%figure6_a
%\includegraphics[width=6 cm, height=3.5 cm]{ARI_hdb_6} \\
%(a) \hspace{5 cm}  (b)
%\caption{Adjusted Rand Index evaluation for 4-cluster partition (a) and 6-cluster partition (b). The axis labels correspond to the selected partitions in Fig.~\ref{fig:4}a, numbered in descending order of SI, in particular, the partition marked by ``$\times$''  correponds to  the label 0 in the panel (b).  The results obtained via HDBSCAN.}\label{fig:6}
%\end{figure}

Such high ARI values indicate a strong similarity between the selected partitions. The closer the ARI is to 1, the fewer differences can be distinguished between the partitions. For instance, when comparing two partitions with the same number of clusters, the differences may manifest in the location of individual points, or small portions of one cluster may overlap with another cluster.

Thus, the ARI evaluation allows us to categorize the selected partitions into two classes: those that are similar to each other and those that are significantly different. In other words, a threshold ARI value (which we have determined) can be introduced to distinguish between partitions that are most similar from an informational perspective. Therefore, to represent the similar partitions, we can select any one of them.

Further reduction in the number of partitions requires additional information about the earthquakes. In particular, we compare the clusters with the structure of the fault network. 
 Moreover, we depicted  earthquakes with magnitudes of M$>$5.5 (filled bullets in Figs. \ref{fig:3}c,d and \ref{fig:4}b,c,d) representing  the strongest earthquakes in this region and to some extent relate to local seismically active zones. For instance, as reported by \cite{Atabekov_strong_quake} regarding the regions in Central Asia (Uzbekistan),
the strong earthquakes (M $\geq$ 5) are often linked to the activation of deep faults within seismically active zones. 
Accordingly, we compare the localization of the obtained clusters with the spatial distribution of strong earthquakes. %These earthquakes also supplement the fault network. 

For the 4-cluster partitions obtained using  DBSCAN, the partition ``db 159'' 
 can be selected as a  representative of a set of similar 4-cluster partitions. This partition, depicted in Fig.\ref{fig:3}c, provides the maximum SI value (Fig.\ref{fig:3}a),   its clusters are fairly equally filled with points, and cluster locations correlate well with fault consolidation and aggregation of strong earthquakes.   Similar analysis allows one to select the partition ``hdb 7'' (Fig.\ref{fig:4}b). The ARI values for the pair ``db 159'' and ``hdb 7'' is   0.2 that tells us about the significant difference between partitions. Thus, depending on the aim of the investigation, we can select  ``db 159'', when the focus of further research is mainly on global issues, whereas ``hdb 7'' is more suitable for analyzing regions where earthquakes form dense, highly localized domains.

Considering in the similar manner  6-cluster partitions, we can select the partitions  ``db 192'' and ``hdb 15''. Each of them is a good representative of its class of similar partitions. They possess high values of SI and  ARI of this pair is 0.48 that means a low degree  of similarity.  
Analyzing the cluster locations of these partitions and the fault network (Fig.\ref{fig:3}d and Fig.\ref{fig:4}d), we see that cluster locations
indeed  differ, especially in the southern and southwestern parts of Azerbaijan.     
In particular, we use the partitions  ``db 192'' and ``hdb 15'' for a comparative analysis of the depth distributions for  Azerbaijani earthquakes. 

\section{Analysis of the earthquake depth distributions} 

Using the evaluated clusters, various statistical properties of local seismicity can be estimated, including the depth distribution of earthquake hypocenters. 
\begin{figure}[h]
\centering
\includegraphics[width=6.3 cm, height=4.6 cm]{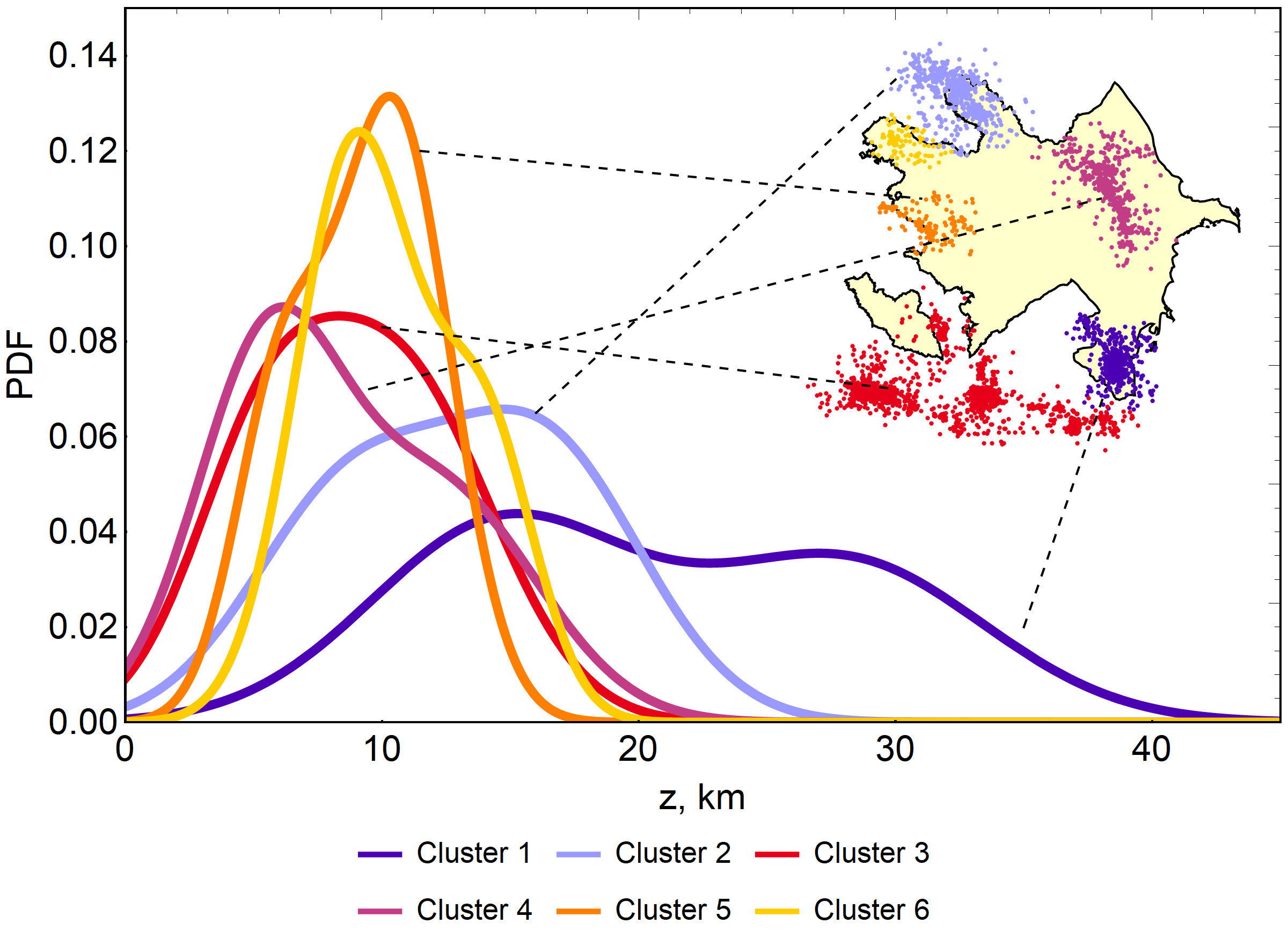}\hspace{0.2 cm}
\includegraphics[width=6.3 cm, height=4.6 cm]{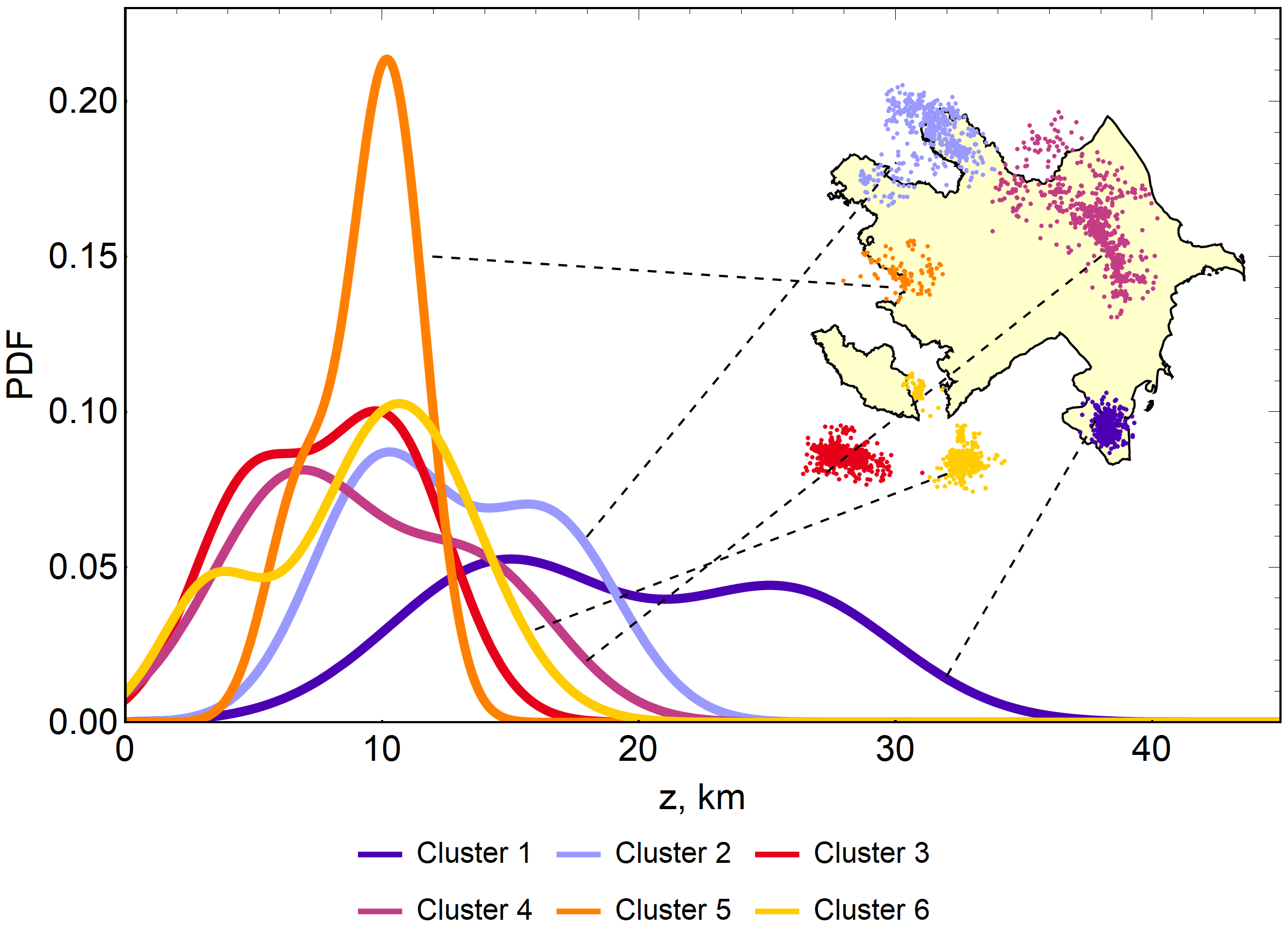} 
\\
(a) DBSCAN, ``db 192'' \hspace{3 cm}  (b) HDBSCAN, ``hdb 15''\\
%\includegraphics[width=6 cm, height=3 cm]{PDF_cluster03}\hspace{0.2 cm}
%\includegraphics[width=6 cm, height=3 cm]{PDF_cluster04} 
%\\
%(c) \hspace{6 cm}  (d)\\
%\includegraphics[width=6 cm, height=3 cm]{PDF_cluster05}\hspace{0.2 cm}
%\includegraphics[width=6 cm, height=3 cm]{PDF_cluster06} 
%\\
%(e) \hspace{6 cm}  (f)
\caption{Fitted PDFs for depths corresponding to the clusters shown in a: Fig.\ref{fig:3}d and b: Fig.\ref{fig:4}d. The color of the curve corresponds to the color of the cluster depicted in the inset.
%Histograms of the earthquake depth  calculated for total number of earthquakes  and  earthquakes belonging to the specified cluster. Solid and dashed curves stand for the approximating PDFs. Clusters are taken from Fig.\ref{fig:4}b.
}\label{fig:6}
\end{figure}

Note that studies of earthquake depth distributions are crucial for understanding the structure of the lithosphere, its rheology, strength, and stress state \cite{Maggi_depth,Jackson_depth}, as well as clarifying the temperature distribution in the Earth's upper layers \cite{Jackson_depth}, assessing the impact of surface activity on local seismicity \cite{Lei_depth}, and more.

As a rule, most earthquakes occur within a single crustal layer \cite{Jackson_depth}, known as the seismogenic crust, which lies above the Moho and is approximately 10 to 40 km thick \cite{Maggi_depth}. Typical histograms of local earthquakes \cite{Pacheco1993depth,Sloan_depth,Nissen_Iran_depth,Jackson_depth} are characterized by unimodal distributions with a pronounced peak at depths ranging from 10 to 20 km. A similar pattern has also been observed in Azerbaijani seismicity, particularly in the Kura fold-and-thrust belt and the surrounding area, as noted by \cite{Tibaldi2024}. However, the depth distribution for the total number of earthquakes worldwide, occurring at depths up to 700 km \cite{VASSILIOU1984depth}, has been found to be bimodal. Furthermore, shallow seismicity observed in subduction zones, such as the Eastern Aleutians, Kermadec Islands, Solomon Islands, and others \cite{Pacheco1993depth}, as well as in the Charlevoix seismic zone (Canada) \cite{Baird_depth} and western Japan \cite{Yano2024}, also exhibits bimodal depth distributions.

Thus,  using the partitions  ``db 192'' and ``hdb 15'', we fit a probability distribution function (PDF) to the histogram representing the depth distribution for each cluster. To do this, we choose PDF in the form of a mixture of two normal distributions, as recommended  by \cite{Pacheco1993depth},  and evaluate the mixture's parameters using the proper Python class, mentioned in Appendix \ref{skur:appPDF}. 
 For comparison, the histogram and corresponding PDF for total earthquake counts are also depicted in Fig.\ref{ds:pdf_approx}. The PDF   exhibits two maxima around 10 km and 32 km, with the upper cutoff depth at 60 km. %and lower cutoff depths for the clusters. 
 This is in good agreement with other seismic research (e.g., \cite{Tibaldi2024}) and the assessment of Moho depth in Azerbaijan, which, according to \cite{Martin_Bochud}, varies from approximately 40 km in the Kura basin and northern foreland basin to over 50 km in the eastern Greater Caucasus.

The application of the fitting procedure to each cluster of the partition ``db 192'' provides the parameters for the PDF curves, presented in Table \ref{skur:Table02_DBS}. The parameters of PDF functions corresponding  to the partition ``hdb 15'' are included in Table \ref{skur:Table02_HDB}. The resulting PDF curves are depicted in   Figs.\ref{fig:6}a and \ref{fig:6}b. In general, the shapes of PDF profiles are mostly similar for the partitions.  In Fig.\ref{fig:6}a, we can distinguish Clusters number 1, which is characterized by the bi-modal PDF. The rest of the clusters are  unimodal. The cluster's PDF maxima are observed at depth 15.31 km and 27.02;  14.72 km; 8.32 km; 6.15 km; 10.28 km; 9.11 km.  The PDF profiles in Fig.\ref{fig:6}b exhibit almost similar properties.  
The distributions associated with clusters 3, 4, and 6 are characterized by bimodality.
%Clusters number 3, 4, and 6 are bi-modal, whereas other ones are unimodal. 
The PDF modes are 15.02 km and 25.1 km;   10.2956 km and 15.668 km;  10.19 km;     6.93 km; 9.74 km; 5.32 km and 10.66 km.
Thus, we see that both algorithms produce four clusters (1,2,3, and 4 in  Fig.\ref{fig:6}), which are nearly identical. This indicates a high degree of reliability in the classification of earthquakes in these areas. Clusters 5 and 6 in Fig.\ref{fig:6}a and the same pair in Fig.\ref{fig:6}b also contain a large number of common points. This means that the algorithms correctly identified seismic events into the clusters. The reasons for the differences in the details of the geometric shape of the clusters may be related to the peculiarities of the algorithms or to the hidden properties of the distribution of earthquakes. The latter circumstance is undoubtedly a particularly intriguing issue for further research.

%As shown by \cite{Pacheco1993depth}, to fit the histograms (see details in \ref{skur:appPDF}), the probability density functions (PDFs) of a mixture of normal distributions are used. 

%The application of this PDF fitting procedure revealed that attempts to approximate the histograms in Figs.\ref{fig:7}c and \ref{fig:7}d using a mixture of normal distributions were unsuccessful. Instead, a single normal distribution was found to be more appropriate. Interestingly, the histograms in Figs.\ref{fig:7}e and \ref{fig:7}f exhibit a bimodal nature. Numerical verification shows that the PDF for the histogram of the total number of earthquake depths has two modes, located at approximately 10 km (intense) and 30 km (almost indistinguishable). The latter peak becomes more pronounced in the cluster histograms (Figs.\ref{fig:7}e, f). In particular, the deeper peak in cluster 6 (Fig.\ref{fig:7}f) is located at a depth of around 35 km. The studies by \cite{Yetirmishli2019} on seismicity in the Ismayilly area, which is covered by cluster 6, also confirm the accumulation of hypocenters at a depth of about 40 km.
A brief note should be made regarding bi-modal PDFs. 
As indicated in \cite{Yano2024}, the two modes observed in the depth histogram for Japanese earthquakes reflect distinct sources: the shallow peak (around 5 km) is associated with crustal earthquakes and anthropogenic events, while the deeper peak (around 35 km) is attributed to seismic events near the plate boundary and deep crustal earthquakes. Other factors contributing to the formation of these two modes may be related to the presence of complex tectonic structures, as discussed in \cite{opentech} and \cite{Yetirmishli2019}. Overall, this issue requires a comprehensive analysis of geological and geophysical data for the region.

\section{Concluding remarks}

In this research, the seismicity of Azerbaijan was studied, with a particular focus on developing a formal strategy for clustering the spatial distribution of earthquake hypocenters. Preliminary evidence of the presence of clusters was obtained by analyzing the Morisita Index, which reliably indicates the potential for cluster selection. Additionally, the Morisita Index evaluation provided an estimate for the fractal dimension of the distributions of both earthquake epicenters and hypocenters.
  
Using the DBSCAN and HDBSCAN algorithms, we evaluated partitions of the earthquake catalog. Since there are no strict restrictions on the intrinsic parameters of the algorithms, we obtained a variety of partitions containing between 2 and 10 clusters. To assess the quality of clustering and select the appropriate partitions for further study, the Silhouette Index was calculated. Partitions providing the maximum Silhouette Index, or those close to it, were selected. The next step in reducing the number of partitions involved using the Adjusted Rand Index. This index was calculated for pairs of partitions with the same number of clusters, allowing us to identify similar partitions. As a result of these clustering stages, we selected partitions containing 4 or 6 clusters, which are the most informative from a data perspective.

Since we performed a formal analysis of the dataset, additional earthquake-related information should be used to make the final partition selection. In this study, we compared the partitions with the regional fault network. This comparison led to the selection of a partition containing 6 clusters, which aligns well with the regional fault network.

Using the selected partition, we studied the earthquake depth distributions by constructing probability density functions (PDFs) for each cluster. To approximate the PDFs, we employed a mixture of normal distributions. It was found that the clusters located along the sea coast are characterized by bimodal distributions.

%\bmhead{Supplementary information}

%If your article has accompanying supplementary file/s please state so here. 

%Authors reporting data from electrophoretic gels and blots should supply the full unprocessed scans for key as part of their Supplementary information. This may be requested by the editorial team/s if it is missing.

%Please refer to Journal-level guidance for any specific requirements.

%\bmhead{Acknowledgements}

\vspace {0.5 cm}
{\bf Funding}
The work is supported in part by  the NATO 
Project G5907 and Project 0121U107662,  National Academy of Sciences of Ukraine, Projects 0123U100183.

\vspace {0.5 cm}

{\bf Data availability} The earthquake catalog used in the current study is available at \url{http://www.isc.ac.uk/iscbulletin/search/}. The fault catalog is available at \url{http://neotec.ginras.ru/index/english/database_eng.html}.
Python source code implementing the cluster analysis of the earthquake catalog is available at \url{https://github.com/SkurativskaKateryna/IJ_NLM_Earthquake_clusterization.git}.

\begin{appendices}

\section{Estimation of  DBSCAN parameters}\label{skur:appA}

\begin{figure}[h]
\centering
\includegraphics[width=6.2 cm,height=4cm]{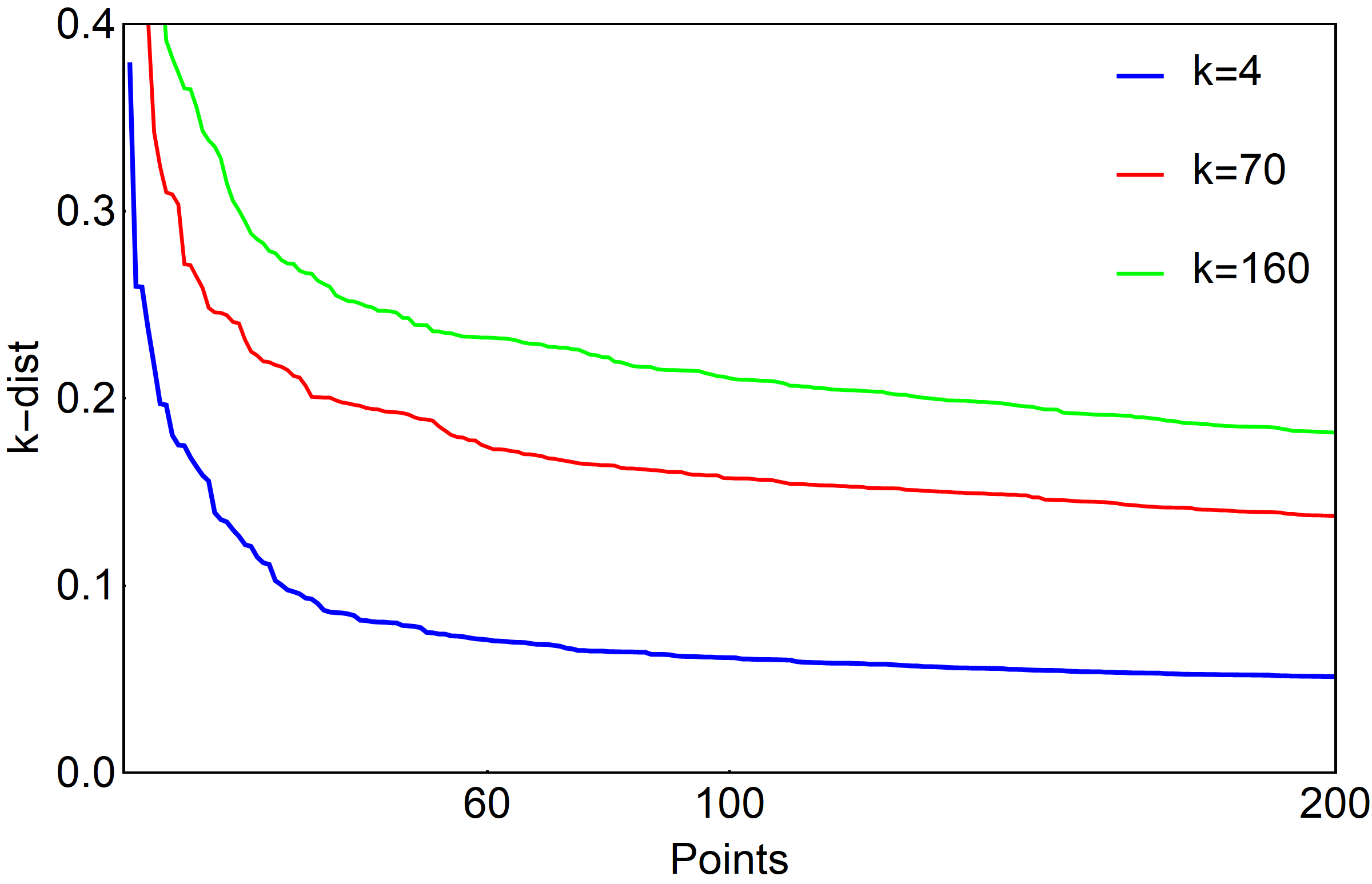}
\caption{The sorted $k$-distance graphs. }\label{ds:kdist}
\end{figure}

The selection of the parameters $\varepsilon$ and $Min\_samples$ for the DBSCAN algorithm is based on the analysis of the sorted $k$-distance graph, as discussed by  \cite{DBSCAN} and \cite{Fana_cluster}, which uses the sorted distance matrix. We plot the $k$-distance graphs for $k=4$, $k=70$, and $k=160$ (Fig.~\ref{ds:kdist}). The appropriate value of $\varepsilon$ corresponds to the formation of the "knee" on these graphs. Therefore, for our  studies, we select $\varepsilon$ from the range $[0.01;0.2]$ and $Min\_samples$ from the range $[10;300]$. Subsequent calculations indicate that the latter range can be narrowed to $[80, 200]$, which is used throughout the paper. 
   It is important to note that this procedure should still be considered an approximate method for parameter estimation, as noted by \cite{Piegari_cluster} and \cite{Cesca2020}.

\section{Fitting a mixture of normal distributions to earthquake depth data}\label{skur:appPDF}

\begin{figure}[tbh]
\centering
\includegraphics[width=6.2 cm,height=4cm]{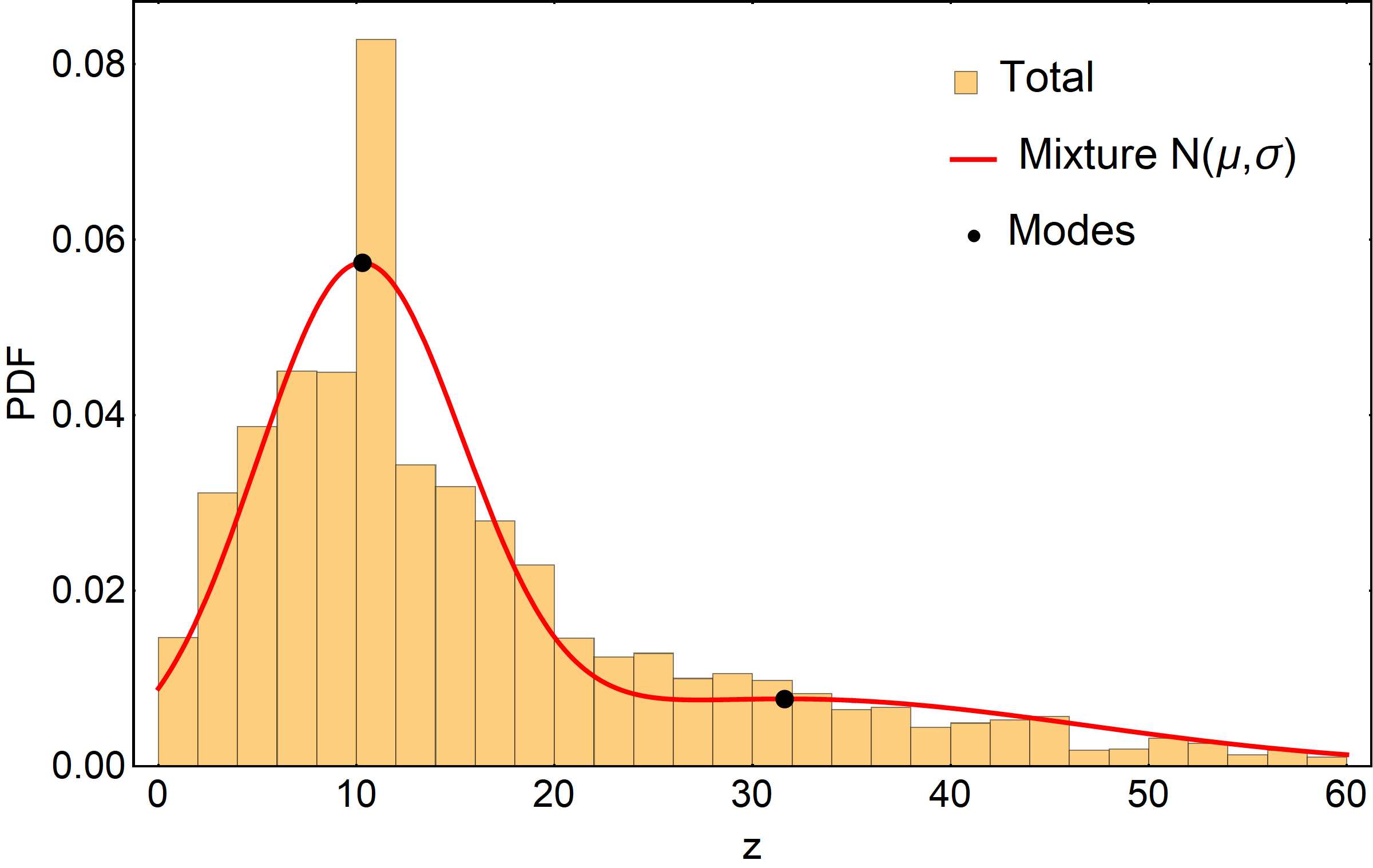}
\caption{Fitting the mixture of normal modes to the earthquake depth histogram. The locations of modes (filled black dots) are 10.29 km and 31.63 km.}\label{ds:pdf_approx}
\end{figure}

To evaluate the PDFs plotted in Fig.\ref{fig:6}, we fit  a mixture of normal distributions to the corresponding depth histograms for each cluster. To do this,  we used the  Python module {\it  sklearn.mixture} containing the class {\it GaussianMixture}  providing  tools for working with Gaussian Mixture Models (GMMs). Particularly, we specify  two component mixture of normals, i.e., $a N_1(\mu_1,\sigma_1)+b  N_2(\mu_2,\sigma_2)$. 
The parameters $a$, $b$  ($a+b=1$), the expectations $\mu_{1,2}$, and the standard deviations $\sigma_{1,2}$ should be estimated. For instance, considering the total number of earthquakes, the depth histogram can be constructed  (Fig.\ref{ds:pdf_approx}), and corresponding parameters of the mixture of normals   $a= 0.7119$,  $\mu_1= 10.1612$, $\sigma_1=  5.1950$, $b=0.2881$, 	$\mu_2= 31.8807$,  and $\sigma_2=  14.99 $ are calculated.

 For the clusters obtained by DBSCAN, the calculated values of parameters are presented in Table \ref{skur:Table02_DBS}. Corresponding PDF profiles are shown in Fig.\ref{fig:6}a.
  Considering  the clusters obtained by HDBSCAN, we evaluated parameter values  presented in Table \ref{skur:Table02_HDB} and plot the corresponding PDF profiles in Fig.\ref{fig:6}b.

 \begin{table}[h]
\caption{The  parameters of the mixture of normal distributions (partition ``db 192'') depicted in Fig.\ref{fig:6}a. }\label{skur:Table02_DBS}
\centering 
\begin{tabular}{@{}lllllll@{}}%{ c|c|c|c|c|c|c } 
 \toprule
  									& $ a$ & $\mu_1$ 			& $\sigma_1$ 	 & $b$ 			& $\mu_2$ & $\sigma_2$  \\  \midrule
  Cluster 1               &0.5403& 14.7111& 5.141&  0.4597& 28.0154& 5.4431 \\ 
Cluster 2 & 0.5580&16.264& 3.8686 & 0.442& 8.6523& 3.7392 \\
Cluster 3  & 0.649 & 10.7328 & 3.596& 0.351& 5.2912 & 2.8054\\
  Cluster 4  & 0.5357& 5.5324 & 2.8294 & 0.4643& 12.2318& 3.7938\\
 Cluster 5  &0.7111 & 10.4547& 2.2046 & 0.2889 & 5.9833 & 1.7462\\ 
 Cluster 6  & 0.7061 & 8.961& 2.3141 & 0.2939& 13.9953& 1.9496\\ 
 \bottomrule
\end{tabular}
  \end{table}
  
  \begin{table}[h]
\caption{The  parameters for the mixture of normal distributions (partition ``hdb 15'') depicted in Fig.\ref{fig:6}b. }\label{skur:Table02_HDB}
\centering 
\begin{tabular}{@{}lllllll@{}}%{ c|c|c|c|c|c|c } 
 \toprule
  									& $ a$ & $\mu_1$ 			& $\sigma_1$ 	 & $b$ 			& $\mu_2$ & $\sigma_2$  \\  \midrule
% Total 	depths		& 0.674  &  0.326 	 	& 9.945 	&  4.842 	& 30.147    &   15.137 \\  \hline
 Cluster 1               & 0.6086  & 6.4897  & 3.1488 & 0.3914 & 13.6962 &3.1542  \\ 
Cluster 2 & 0.7563  & 10.2446 &  1.4319 &   0.2437  & 6.8023 &1.3428 \\
Cluster 3  & 0.5452  & 9.91  & 2.6428 &    0.4548  & 16.4508 & 2.7848 \\
  Cluster 4  & 0.5604 & 14.6678 & 4.3653 &   0.4396 & 25.7917 & 4.1998  \\
 Cluster 5  &0.5915 & 10.1408 &2.4734 &
   0.4085 & 4.7844 & 2.2276\\ 
 Cluster 6  &0.8044 & 10.6700 & 3.1267 &
   0.1956  & 3.2650 & 1.9019\\ 
 \bottomrule
\end{tabular}
  \end{table}
  
%  \begin{table}[tbh]
%\caption{The  parameters for the mixture of normal distributions depicted in Fig.\ref{fig:7}. }\label{skur:Table02}
%\begin{tabular}{@{}lllllll@{}}%{ c|c|c|c|c|c|c } 
% \toprule
%  									& $ a$ & $b$ 		& $\mu_1$ 			& $\sigma_1$ 		& $\mu_2$ & $\sigma_2$  \\  \midrule
% Total 	depths		& 0.674  &  0.326 	 	& 9.945 	&  4.842 	& 30.147    &   15.137 \\  \hline
 %Fig.\ref{fig:7}(a)  & 0.644 & 0.356 	&11.148 & 5.011	&24.235 & 10.126 \\ 
 %Fig.\ref{fig:7}(b)  & 0.937 & 0.063 & 8.140 & 3.749 & 21.719 & 9.272\\
 % Fig.\ref{fig:7}(c)  & -- & -- & 9.892 & 4.291 & -- & --  \\
 %  Fig.\ref{fig:7}(d)  &   -- & -- & 9.614 & 4.986& -- & --  \\
% Fig.\ref{fig:7}(e)  &0.698 & 0.302 & 2.544 & 4.362 & 8.444 & 3.223\\ 
 %Fig.\ref{fig:7}(f)  & 0.732 & 0.268 & 9.166 & 4.849 & 34.938 & 10.027\\ 
% \botrule
%\end{tabular}
%\end{center}

%\end{table}
  
%It turned out that this procedure can fail when  $b$ is close to zero. Then  we applied {\sf EstimatedDistribution[data, NormalDistribution[$\mu_1$, $\sigma_1$]}.

\end{appendices}


%\bibliography{cluster3D_skur}% common bib file

%\end{document}

\bibliographystyle{IEEEtran_link}

\bibliography{cluster3D_skur}% common bib file

\end{document}